\documentstyle[preprint,aps,pre]{revtex}
\tighten
\begin{document}
\draft
\title{Critical behaviour of the Random--Bond Ashkin--Teller Model
       --- a Monte--Carlo study.}
\author{Shai Wiseman\cite{email1} and Eytan Domany\cite{email2}}
\address{Department of Physics of Complex Systems, Weizmann Institute of
science, \\          Rehovot 76100 Israel }
\date{October 19, 1994}
\newcommand{\av}[1]{\langle #1 \rangle}
\maketitle
\begin{abstract}

The critical behaviour of a bond-disordered Ashkin-Teller model on a square
lattice is investigated by intensive Monte-Carlo simulations. A duality
transformation is used to locate a critical plane of the disordered model.
This critical plane corresponds to the line of critical points of the pure
model, along which critical exponents vary continuously. Along this line the
scaling exponent corresponding to randomness $\phi=(\alpha/\nu)$ varies
continuously and is positive so that randomness is relevant and different
critical behaviour is expected for the disordered model. We use a cluster
algorithm for the Monte Carlo simulations based on the Wolff embedding
idea, and perform a finite size scaling study of several critical models,
extrapolating between the critical bond-disordered Ising and
bond-disordered four state Potts models. The critical behaviour of the
disordered model is compared with the critical behaviour of an anisotropic
Ashkin-Teller model which is used as a refference pure model.
We find no essential change in the order parameters' critical exponents with
respect to those of the pure model. The divergence of the specific heat $C$
is changed dramatically. Our results favor a logarithmic type divergence at
$T_{c}$, $C\sim \log L$  for the random bond Ashkin-Teller and four state
Potts models and $C\sim \log \log L$ for the random bond Ising model.

\end{abstract}

% =-=-=-=-=-=-=-=-=-=-=-=-=-=-=-=-=-=-=-=-=-=-=-=-=-=-=-=-=-=-=-=-=-=-=-=-=
\narrowtext

\section{Introduction}
 How is the critical behaviour affected by the introduction of
disorder ( usualy dilution or bond-randomness ) into a model ?
 The Harris criterion
 \cite{Harris,Stinch,Kinz:Dom}
 states that $\phi$, the scaling index of the operator
corresponding to randomness at the pure system fixed point
(also called the crossover exponent ) is equal to
$\frac{\alpha}{\nu}$ ($ \alpha$ and $\nu$ are the specific-heat and
correlation length exponents of the pure model). Thus the critical behaviour
of the pure system is unaltered by disorder if $\alpha<0$, and altered when
$\alpha>0$. If $\alpha=0$ the situation is marginal.

Renormalization-group methods (namely expansion in the
parameter $\epsilon=4-d$, where $d$ is the dimensionality), applied to
n-component continuous
spin models with weak quenched bond disorder\cite{L and H}, confirmed the
Harris criterion $\phi=\frac{\alpha}{\nu}$. Moreover, it was found that for
$n<n_{c}=4-4\epsilon+O(\epsilon^{2})$ a new stable `random' fixed point
with new exponents appears.
For the case $n=1$ (Ising model) the stability of an $O(\epsilon^{1/2})$
 random fixed point was shown by Jayaprakash and Katz\cite{J and K}.
For the $d=2$ Ising model $\alpha=0$, and the operator corresponding
to randomness is expected to be marginally relevant.

 Two partially conflicting theories by Dotsenko and Dotsenko\cite{Dots:Ising}
(DD) and
by Shalaev\cite{Shal}, Shankar\cite{Shan}, and Ludwig\cite{Lud} (SSL) were
suggested for the $d=2$ Ising
model with small randomness. While both theories agreed on the
divergence of the specific heat, e.g.
\begin{equation}
 C_{\text{imp}}(t)\sim \ln \ln 1/|t|  \;\;\;\;,
\label{eq: C(t)ising ran} \end{equation}
conflicting predictions were made for other quantities like the
magnetization $ M$ and susceptibility
$\chi$.   Extensive Monte-Carlo (MC) simulations of dilute and random-bond
Ising models ( see a review in ref. \onlinecite{rev selke} and references
therein) helped to confirm
the prediction on   the specific heat. The other quantities were seen to
behave ( as predicted by SSL), in essence, as in the pure
model. The disordered $d=3$ Ising model has been studied by means of MC
simulations as well\cite{rev selke}.

 Less attention has been paid in recent times to other disordered
systems such as  random q-state Potts models.
 Kinzel and Domany \cite{Kinz:Dom} used a real-space
 Migdal-Kadanoff renormalization transformation for the random-bond Potts
model on the square lattice in two dimensions and found ( when
$\alpha_{pure}>0$) a stable random
fixed point with a reduced (with respect to the pure model) but positive
$\alpha_{random}$. Using methods of the same spirit, Andelman and Berker
\cite{Andelman} also found a random fixed point. For $q=4$ their calculation
suggests $\alpha=-.37$ and $\nu=1.19$.
Other workers\cite{Der Gar,Lud Q1} too have suggested a negative $\alpha$
for $q>2$ Potts models in two dimensions.
MC work on random q-state Potts
models has not been carried out, except for recent work on the bond-random
two dimensional 8-state Potts model\cite{8potts}, where randomness was shown
to change
the first order phase transition of the pure model into a second order
one\cite{Berker}.
 Novotny and Landau (NL)\cite{Nov Lan} used MC simulations to study the two
dimensional site-dilute Baxter-Wu\cite{Baxter Wu} model which is of the same
universality class as the 4-state Potts model. NL measured the exponents
$\alpha=0.0(2)$, $\nu=1.00(7)$, and $\gamma=1.95(8)$  which did not seem to
depend on the amount of dilution. They could not determine whether the new
stable fixed point is a $d=2$ random Ising fixed point (which is very
similar to the pure Ising fixed point) or a new impure Baxter-Wu fixed
point.

In this paper we address the effect of randomness on a
critical line which connects the critical points of the $d=2$ Ising and
4-state Potts models. Such is the critical line of the two
dimensional Ashkin-Teller model\cite{AT:1} on a square lattice.

A convenient representation of the pure Ashkin-Teller (AT) model is
in terms of two Ising spin variables, $\sigma_i$ and $\tau_i$, placed
on every site of a lattice. Denoting by $<ij>$ a pair of nearest
neighbor sites, the Hamiltonian is given by
\begin{equation}
{\cal H}=-\sum_{<ij>} [ K \sigma_i\sigma_j + K \tau_i \tau_j +
\Lambda\sigma_i\sigma_j  \tau_i \tau_j]
  \label{eq:AT1} \end{equation}
Here $K$ is the strength of the interactions
between neighboring $\sigma$ and $\tau$ spins, and $\Lambda$ is a four-spin
coupling ( throughout this article we absorb a factor of $1/k_{B}T$ into
the coupling constants).
The phase diagram of the ferromagnetic AT model is known in two dimensions
from duality transformations and renormalization group
studies\cite{Wu:Lin,Domany:Riedel}.  The three dimensional model has
been studied as well\cite{AT3d}.
 The phase diagram which we reviewed previously\cite{SE} includes
a line of critical points which connects the Ising critical point at one
end to the 4-state Potts critical point at the other end. Along this
line the exponents
vary continuously, and have been determined analytically
\cite{exact,Nienhuis},
interpolating between their Ising and four-state Potts values.
For instance,
%the value of the ratio $\alpha/\nu$
%varies from 0 at the Ising ($\Lambda=0$) critical point to
%$\alpha/\nu=1$ at the four-state Potts point $K=\Lambda$.
 the crossover exponent connected with randomness $\phi=\frac{\alpha}{\nu}$
changes smoothly from 0 at the decoupled Ising point, to 1 at the
4-state Potts point. Therefore, according to the Harris criterion, we expect
randomness to be a relevant
operator of varying strength in this regime of the AT model, and expect
 the critical behaviour of the disordered model to differ from that of the
pure model.

  The critical behaviour of the disordered AT model has not been
studied before,
 apart from a conjecture by Alcaraz and Tsallis \cite{Alcaraz} as to the
location of the critical manifold of the bond dilute AT model.
 None the less the effect of disorder on the critical line of the
Baxter( or symmetric 8-vertex) model\cite{Baxter}, which is isomorphic to
the critical line of the AT model\cite{exact,Nienhuis}, was considered.
DD\cite{DD:bax} have extended their study of the disordered Ising model
to the Baxter model,and found, for small disorder, the specific heat
divergence
to change from its pure form to the form of eq. (\ref{eq: C(t)ising ran}).
Matthews-Morgan, Landau and Swendsen  (MLS)\cite{MLS} have studied the
site-dilute Baxter model by means of a MCRG ( MC renormalization-group)
method. MLS found
$\frac{1}{\nu}=y_{T}=0.98(7)$ and $\frac{\gamma}{\nu}$ slightly below the
Ising value.
 Both studies indicate that continuous variation of
critical exponents of the pure model is substituted by flow to a random
 Ising fixed point.

In a previous article\cite{SE} we described an efficient MC cluster
algorithm\cite{rev:clust}
for the AT model. In this work we have used this algorithm to perform
extensive MC simulations\cite{BH:book} of the random-bond AT model. The
importance
of using a cluster algorithm for the acquirement of reasonably accurate
data cannot be overlooked. This is especially true in the highly random
regime
where local algorithms are extremely slow\cite{Heuer} while cluster
algorithms are most efficient\cite{Wang1,Hen}.The partial elimination of
critical slowing down allows us to go to large lattice sizes (up to
L=256) and obtain
accurate results on the finite size scaling of the thermodynamic quantities
at criticality. Our results seem to favor an Ising like critical behaviour.
% with those of DD for the Baxter model.

This work is organized as follows. In sec. \ref{sec:RBAT} we define the
random-bond AT model and a related anisotropic AT model. As will be
explained, the anisotropic AT model is used as a non-disordered reference
model for the random-bond model. We use the duality
transformation of the AT model to locate exactly critical surfaces of the
two models, which are related to the line of critical points of the pure
model. In sec. \ref{sec:details} we describe the MC cluster algorithm, give
definitions of the measured quantities and describe our choice of simulation
points.We have performed simulations of
the anisotropic and random-bond AT models at several points on their critical
surface. In sec. \ref{sec:results} we give our results which consist of the
finite size dependence of thermodynamic quantities of the two models at
criticality. Analysis of our results shows that the anisotropic
model exhibits varying critical exponents, i.e. the specific heat
 diverges as
\begin{equation}
 C(t)\sim |t|^{-\alpha}  \;\;\;\;,
\label{eq: C(t)ani} \end{equation}
 where at different points of the critical surface of the
anisotropic AT model we measured different values of $\alpha$.
 On the other hand,
 the random-bond model
seems to exhibit a single type of
behaviour, i.e. the specific heat diverges as
\begin{equation}
 C_{\text{imp}}(t)\sim - \ln |t|  \;\;\;\;.
\label{eq: C(t)ising} \end{equation}
 this is identical with a pure Ising behaviour and not with a random Ising
one.
% accord with the prediction of DD for the Baxter model.
In sec. \ref{sum} we summarize our results.

\section{Definition of Models and Location of Critical Planes }
\label{sec:RBAT}
The model we wish to study is the {\em Random-Bond} AT model (RBAT) on a
square lattice, which is defined by the Hamiltonian
\begin{equation}
{\cal H}=-\sum_{<i,j>}[K_{i,j}\sigma_{i}\sigma_{j}+K_{i,j}\tau_{i}\tau_{j}+
\Lambda_{i,j}\sigma_{i}\tau_{i}\sigma_{j}\tau_{j}] \;.
   \label{eq:rbAT}  \end{equation}
 The coupling constants $K_{i,j}$ and $\Lambda_{i,j}$
 are chosen according to
   \begin{equation}
(K_{i,j},\Lambda_{i,j})=\left\{ \begin{array}{ll}
 \,(K^{1},\Lambda^{1}) & \mbox{with probability } p \\
 \,(K^{2},\Lambda^{2}) & \mbox{with probability } 1-p \end{array}
\right. \label{eq:Ran} \;.   \end{equation}

 It seems quite natural ( the {\em need} for this comparison will be made
clear at the end of this section) to compare the RBAT with the {\em
Anisotropic}
AT model (AAT)  with the same Hamiltonian (\ref{eq:rbAT}) but with the
couplings  distributed as follows:
   \begin{equation}
(K_{i,j},\Lambda_{i,j})=\left\{ \begin{array}{ll}
 \,(K^{1},\Lambda^{1}) & \mbox{ for bonds $(i,j)$ in the
horizontal direction } \\
 \,(K^{2},\Lambda^{2}) & \mbox{for bonds $(i,j)$ in the vertical direction }
\end{array} \right. \label{eq:Ani} \;.   \end{equation}

Spatial anisotropy is usualy known to be an irrelevant operator, that is,
the critical behaviour of a model is unaltered by introduction of
anisotropy. Thus one expects that continuous variation of critical exponents
will continue to exist on some critical manifold of the AAT, as is the case
for the isotropic (or pure) model.

A duality transformation can be used to locate critical manifolds of the
AAT and RBAT in a way we now present.

\subsection{Duality and location of a phase transition }\label
 {sec:duality ani}
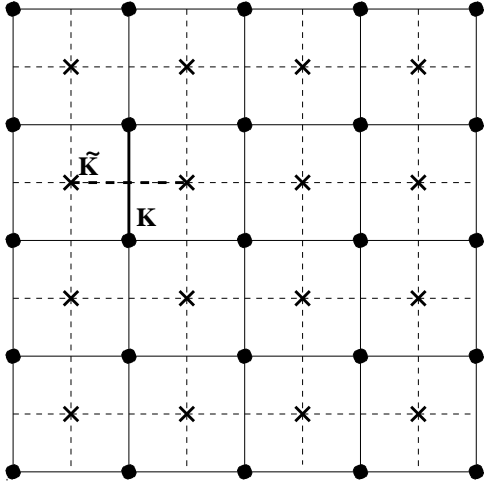
\begin{figure}[h]
\caption{The original lattice $G$ is denoted by solid lines, while its sites
are denoted by dots. The dual lattice $D$ is denoted by dashed lines, while
its sites are denoted by crosses. A bond of strength $K$ is denoted by a
thick solid line, while its dual of strength $\tilde{K}$ is denoted by a
thick dashed line.}
\label{fig:duality exp}
\end{figure}

 Consider an AT system on a square lattice with spatially varying coupling
constants (the spatial variation can be due to anisotropy, or it can be due
to randomness as we consider later). The dual lattice $D$ of the original
square lattice $G$ is the square
lattice whose sites are at the centers of the plaquettes of the original
lattice. In figure \ref{fig:duality exp} sites of $G$ are denoted by dots,
while sites of $D$ are denoted by crosses. Consider a bond of
strength $(K,\Lambda)$ between two sites in $G$. Under duality it is mapped
onto a bond of strength
$(\widetilde{K},\widetilde{\Lambda})=\vec{D}[(K,\Lambda)]$,
where $\vec{D}$ is the duality transformation of the AT
model\cite{Wu:Wang,Domany:Riedel} which we list in the Appendix.
The dual bond lies in the dual lattice $D$, intersecting the original bond.
This situation is depicted in figure \ref{fig:duality exp}. The original
bond of strength $(K,\Lambda)$ is depicted by the solid line while its dual
is depicted by the dashed line.

Suppose the only spatial variation of the couplings is one of anisotropy.
 Then an AT system with couplings of strength $(K^{1},\Lambda^{1})$
in the horizontal direction and strength $(K^{2},\Lambda^{2})$ in the
vertical
direction, transforms under duality into an AT system on the dual lattice
$D$ with couplings of strength $(\widetilde{K^{1}},\widetilde{\Lambda^{1}})$ in
the
vertical direction and strength $(\widetilde{K^{2}},\widetilde{\Lambda^{2}})$
in the
horizontal
direction. The free energies of the original system and its dual  have the
same singular part, and therefore we can write
\begin{equation}
f_{s}[(K^{1},\Lambda^{1}),(K^{2},\Lambda^{2})]=
f_{s}[(\widetilde{K^{2}},\widetilde{\Lambda^{2}}),(\widetilde{K^{1}},
\widetilde{\Lambda^{1}}) ] \;,
\label{eq:fs Anidual}  \end{equation}
where the first pair of couplings corresponds to the horizontal bonds,
and the second pair to the vertical ones. The duality transformation
$\vec{D}$ is self-inverse. Thus,
an anisotropic AT system with couplings
$[(K^{1},\Lambda^{1}),(K^{2},\Lambda^{2})]$ such that
$(K^{2},\Lambda^{2})=(\widetilde{K^{1}},\widetilde{\Lambda^{1}})$, transforms
under
duality onto itself. So,
there exists a two dimensional self-dual manifold (it would help the reader
 to view the duality transformation as a mathematical mapping in the four
dimensional space of couplings $[(K^{1},\Lambda^{1}),(K^{2},\Lambda^{2})]$ ).

 The line of critical points of the isotropic AT model lies within the
 subspace $\Sigma$, defined by $0 \leq \Lambda \leq K$. $\Sigma$ is located
between a decoupled Ising
model where $\Lambda=0$ and a four-state Potts model with  $\Lambda=K$. Two
sub-areas of $\Sigma$
 which map onto each other under duality are separated by a self dual
line ( fulfilling $(K,\Lambda)=(\widetilde{K},\widetilde{\Lambda})$ ).
 The assumption that $\Sigma$
 contains only two phases, with a single phase transition between
them(where $f_{s}$ is non-analytic) leads to the conclusion that this phase
transition must occur on the self-dual line.
In this way the line of critical points of the isotropic AT
was located exactly\cite{Wu:Lin}.

Let us now consider the corresponding subspace, $\Sigma_{A}$, of the AAT for
which $0 \leq \Lambda^{1} \leq K^{1}$ and $0 \leq \Lambda^{2} \leq K^{2}$.
$\Sigma_{A}$ lies between an anisotropic decoupled Ising subspace where
$\Lambda^{1}=\Lambda^{2}=0$ and an anisotropic four-state Potts subspace
with $\Lambda^{1} = K^{1},\;\Lambda^{2} = K^{2}$. $\Sigma_{A}$ is invariant
under duality since for  $0 \leq \Lambda \leq
K \;$ we get (see Appendix)  $\;0 \leq \widetilde{\Lambda} \leq \widetilde{K}$.
 It follows also that the self dual plane $0 \leq \Lambda^{1} \leq K^{1}$ ,
$(K^{2},\Lambda^{2})=(\widetilde{K^{1}},\widetilde{\Lambda^{1}})$ is
included in the subspace we are considering. Thus, again assuming that
$\Sigma_{A}$  contains only two phases, with a single phase transition
between
them, we conjecture that the self dual plane is a plane of phase transitions
(this kind of
argument was first used by Fisch\cite{Fisch} for the random-bond Ising
model, and later for Potts models by Kinzel and Domany\cite{Kinz:Dom}).
 Since the space of couplings is four dimensional, while the self dual plane
is only
two dimensional, the self dual plane is only {\em part} of the three
dimensional critical manifold in $\Sigma_{A}$ . Notice that this
is unlike the case of the Potts model or any model with a single coupling
constant where the self dual line is {\em identical} with the critical line.

\subsection{A Critical Plane of the RBAT } \label{sec:duality
ran}
 The argument that leads to locating the corresponding critical
plane of the RBAT is very similar to the one presented above, up to a subtle
difference.
Consider a specific realization of a RBAT on a square lattice with couplings
$(K^{1},\Lambda^{1}),\,(K^{2},\Lambda^{2})$
distributed at random with probabilities $p$
and $1-p$, respectively. Replacing each bond by its dual we get (on the dual
lattice) a system with couplings $(\widetilde{K^{1}},\widetilde{\Lambda^{1}})$
and
$(\widetilde{K^{2}},\widetilde{\Lambda^{2}})$, distributed at random with
probabilities $p$ and
$1-p$. This defines a one to one mapping under duality from the ensemble of
realizations of $[(K^{1},\Lambda^{1}),\,(K^{2},\Lambda^{2}),p]$
RBAT systems, onto the
ensemble of the dual systems
$[(\widetilde{K^{1}},\widetilde{\Lambda^{1}}),(\widetilde{K^{2}},\widetilde{\Lambda^{2}}),p
]$. Under duality each bond is mapped onto the bond on the dual lattice
which intersects the original bond (see fig. \ref{fig:duality exp}). Thus,
the spatial distribution of the two types of bonds on the dual lattice is
different from that of the original lattice. But since all spatial
distributions have the same weight in their ensemble, there is a one to
one correspondence of bond configurations in the two ensembles, and we can
write:
\begin{equation}
f_{s}[(K^{1},\Lambda^{1}),(K^{2},\Lambda^{2}),p]=
f_{s}[(\widetilde{K^{1}},\widetilde{\Lambda^{1}}),(\widetilde{K^{2}},
\widetilde{\Lambda^{2}}),p ] \;,
 \label{eq:fs rbdual}  \end{equation}
where it is understood that in calculating $f_{s}$ we average over all
spatial bond distributions.
 Choosing $p=\frac{1}{2}$ we find that there is a self dual manifold
 at $(K^{2},\Lambda^{2})=(\widetilde{K^{1}},\widetilde{\Lambda^{1}})$. Our
ability
to locate a self dual manifold for the case $p=\frac{1}{2}$ makes this
choice of $p$ a convenient case for studying the RBAT.
Viewing the duality transformation as a mathematical mapping
in the space of couplings
$[(K^{1},\Lambda^{1}),\,(K^{2},\Lambda^{2}),\frac{1}{2}]$,
it is identical to the duality transformation in
the space of couplings $[(K^{1},\Lambda^{1}),\,(K^{2},\Lambda^{2})]$ of the
anisotropic
AT model. Thus all conclusions which were based on duality in sec.
\ref{sec:duality ani} apply here equally.
Following the logic of the previous subsection we conjecture that the
subspace of the RBAT for which $0 \leq \Lambda^{1} \leq K^{1}$ , $0 \leq
\Lambda^{2} \leq K^{2}$ and $p=\frac{1}{2}$ includes in it a critical plane
defined by $0 \leq \Lambda^{1} \leq K^{1}$ ,
$(K^{2},\Lambda^{2})=(\widetilde{K^{1}},\widetilde{\Lambda^{1}})$ .

In order to study the effect of randomness on the AT model we need a
reference non-disordered system for comparison.
When studying a specific RBAT model with couplings
$[(K^{1},\Lambda^{1}),\,(K^{2},\Lambda^{2}),\frac{1}{2}]$ it is not clear
with which specific non-random AT model with couplings $(K,\Lambda)$ should
one compare.
 None the
less a comparison with the corresponding AAT model with couplings
$[(K^{1},\Lambda^{1}),\,(K^{2},\Lambda^{2})]$ seems to be quite natural.
Consider the anisotropic model with bonds $(K^{1},\Lambda^{1})$ in the
horizontal direction and bonds $(K^{2},\Lambda^{2})$ in the vertical
direction.
Next, consider as a perturbation, changing a small fraction $p$ of the bonds
in the horizontal direction into bonds of strength $(K^{2},\Lambda^{2})$ and
the same fraction of bonds in the vertical direction into bonds of strength
 $(K^{1},\Lambda^{1})$. When $p$ increases all the way to $p=\frac{1}{2}$
the anisotropy disappears and the
model  is completely random. So, in a sense, the perturbation of randomness
can take us continuously from the critical plane of the anisotropic model
to the critical plane of the random model.
In view of this simple picture we have performed simulations of the
anisotropic and random-bond AT models at several points on their self-dual
planes described above. We indeed find the two planes to be critical.
In the next section we give some details of the simulations.

%More precisely consider a critical model of the AAT described by
%$[(K^{1},\Lambda^{1}),\,(K^{2},\Lambda^{2})]$ and suppose that under a
%renormalization group transformation it flows to some fixed point $F$.
%Under the
%According to the Harris
%criterion, we expect this perturbation of randomness to be relevant, and
%expect to find different critical behaviour for the random model.
%All our
%measurements were performed at this self-dual plane of the RBAT and at the
%corresponding self-dual plane of the  AAT. We indeed find the two planes to
%be critical.

\section{Method and Details of Simulations} \label{sec:details}
\subsection{ The Method }  \label{sec:method}
In ref. \onlinecite{SE} we described a cluster algorithm for the AT model and
tested its performance (on the isotropic model). as explained there, the
algorithm can be regarded from two different points of view. The
conceptually simpler
and faster ( though less general ) one is that of the Wolff embedding
idea\cite{Wol:1C,embed:1}. It is faster in the sense
that its implementation on the computer runs faster since the freeze-delete
procedure is less complicated.
 The main idea is to embed into the AT
model an Ising model of space dependent couplings $J_{jk}$ and simulate it
 using the SW\cite{SW:1} or Wolff\cite{Wol:1C} procedure for the Ising
model. To be explicit,
consider the Hamiltonian (\ref{eq:rbAT}), and take the $\tau$ variables as
fixed, so we can write
   \begin{equation}
{\cal H}={\cal H}_{1}+{\cal H}_{2}=-\sum_{<i,j>}(K_{i,j}+
\Lambda_{i,j}\tau_{i}\tau_{j})\sigma_{i}\sigma_{j}-\sum_{<i,j>}K_{i,j}
\tau_{i}\tau_{j} .    \label{eq:AT-Is}  \end{equation}
${\cal H}_{2}$ represented by the second sum is a constant, and remembering
that we are considering  $\Lambda_{i,j} \leq K_{i,j}$ for all (i,j),
 \ ${\cal H}_{1}$ is a ferromagnetic Ising model in the $\sigma$
variables with couplings $J_{i,j}=K_{i,j}+ \Lambda_{i,j}\tau_{i}\tau_{j}$.
Simulating this Ising model with any procedure that will maintain
detailed balance with respect to ${\cal H}_{1}$ and will not change the
value of ${\cal H}_{2}$, will also maintain
detailed balance with respect to ${\cal H}$. So we can use for example the
SW\cite{SW:1} or Wolff\cite{Wol:1C} procedure for the Ising model.
 This by itself will maintain detailed balance but
will not be ergodic since the $\tau$ variables will not be updated.
Obviously to update the $\tau$ variables the same process should be repeated
, holding the $\sigma$ variables fixed and simulating an Ising Hamiltonian
with the $\tau$ variables. To summarize, our procedure goes as
follows: Choose at random whether to embed into the AT Hamiltonian an
Ising Hamiltonian in the $\sigma$ (or $\tau$) spins. Pick a random
site in the lattice, grow a cluster of $\sigma$ (or $\tau$) spins, using
the Wolff single cluster procedure with the Ising Hamiltonian (
(\ref{eq:AT-Is})
or the opposite one in case of a $\tau$ embedding), and flip it.

\subsection{ Choice of Simulation Points}     \label{sec:points}

 All our simulations were performed at the critical plane defined by
\begin{equation}
\{ [(K^{1},\Lambda^{1}),\,(K^{2},\Lambda^{2}),p=\frac{1}{2}] :\;
(K^{2},\Lambda^{2})=(\widetilde{K^{1}},\widetilde{\Lambda^{1}}) \;\;,\;\;
0 \leq \Lambda^{1} \leq K^{1} \;\}  \;,
 \label{eq:def RBS}  \end{equation}
 as explained in sec. \ref{sec:RBAT} (for the anisotropic model just omit
 the $p=\frac{1}{2}$).
For convenience we also use the following notation for the coupling
constants:
\begin{equation}   \begin{array}{lll}
 Z=\exp^{-4K}\; ;\;\;\;\;   & X=\exp^{-2(K+\Lambda)} \; ;\;\;\;\;&
\vec{X}=(X,Z)\;. \end{array} \label{eq:def xz} \end{equation}

The first series of measurements were performed at five points
$(\vec{X^{1}}_{i},\vec{X^{2}}_{i}),\;i=0\ldots4$, which we label as
$C_{i},\;i=0\ldots4$. These were chosen so as to interpolate between
$(\vec{X^{1}}_{0},\vec{X^{2}}_{0})$, which is an anisotropic (or
random-bond) decoupled Ising critical point, and
$(\vec{X^{1}}_{4},\vec{X^{2}}_{4})$,
which is an anisotropic (or random-bond) four-state Potts critical point.
The points $C_{i}$ interpolate in a similar manner to the way in which the
critical
line of the pure AT connects the pure decoupled Ising critical point with the
pure four-state Potts critical point.
Thus we chose the five points $\vec{X^{1}}_{i}$ to lie equidistantly on the
line \begin{equation}
 Z^{1}-Z^{1}_{\text{Ising}}=-2(X^{1}-X^{1}_{\text{Ising}})\;.
\label{eq:line} \end{equation}
 The slope $-2$ was chosen so that the line (\ref{eq:line}) lies
 parallel (in the $XZ$ plane) to the critical line of the pure AT
model, $Z=1-2X$. Note that once $(X^{1},Z^{1})$ are defined, $(X^{2},Z^{2})$
 are defined through (\ref{eq:def RBS}, \ref{eq:def xz}).

We clearly have some freedom in our choice of
$(X^{1}_{\text{Ising}},Z^{1}_{\text{Ising}})$, the decoupled Ising
couplings; setting $\Lambda^{1}=0$ in (\ref{eq:def RBS}) one gets
$\Lambda^{2}=\widetilde{\Lambda^{1}}=0$ as well. The remaining parameter,
$K^{1}$, is chosen so that ( see (\ref{eq:def RBS}) )

\begin{equation}
 K^{2}_{\text{Ising}}=\widetilde{ K^{1}_{\text{Ising}} }=\frac{1}{10}
 K^{1}_{\text{Ising}} .
\label{eq:ratio1} \end{equation}
Since the extent of deviation from pure behavior
 is obviously determined not only by $p$ but also by the
difference between the two sets of couplings, the ratio of $\frac{1}{10}$
was chosen so that randomness will be pronounced\cite{Wang1,r=.5}.

While (\ref{eq:ratio1}) defines the first simulation point
$\vec{X^{1}}_{0}\equiv \vec{X^{1}}_{\text{Ising}}$, the fifth simulation
point $\vec{X^{1}}_{4}\equiv \vec{X^{1}}_{\text{Potts}}$ is chosen to be a
four-state Potts point lying at the intersection of the line (\ref{eq:line})
with the
four-state Potts line $Z^{1}=X^{1}$ (or $\Lambda^{1}=K^{1}$).

The other three simulation points $\vec{X^{1}}_{1,2,3}$ lie equidistantly on
the line (\ref{eq:line}) between $\vec{X^{1}}_{\text{Ising}}$ and
$\vec{X^{1}}_{\text{Potts}}$ :
\begin{equation}   \begin{array}{lll}
X^{1}_{i}=\frac{i}{4}(X^{1}_{\text{Potts}}- X^{1}_{\text{Ising}})+
X^{1}_{\text{Ising}} &  \;\;\;\;
Z^{1}_{i}=Z^{1}_{\text{Ising}}-2(X^{1}_{i}-X^{1}_{\text{Ising}}) &
\;\;\;\;\; i=0\ldots4
\;, \end{array} \label{eq:X1place} \end{equation}
Thus  the first series of points $C_{i}$ at which we made measurements
$C_{0\ldots4}$ is defined through equations
(\ref{eq:def RBS},\ref{eq:def xz},\ref{eq:ratio1},\ref{eq:X1place}) and the
definition of $\vec{X^{1}}_{\text{Potts}}$.

Two additional measurement points were intended to monitor the effect
of varying the amount of randomness in the model on the critical behaviour.
 In order to define these two points we define two series of five
points each, $A_{0\ldots4}$ and $B_{0\ldots4}$ . Similarly to the
series $C_{0\ldots4}$, these two series are defined through equations
(\ref{eq:def RBS},\ref{eq:def xz},\ref{eq:X1place}). But in contrast with
the series $C_{0\ldots4}$, eq. (\ref{eq:ratio1}) is substituted by
\begin{equation}
\frac{ K^{2}_{\text{Ising}} }{ K^{1}_{\text{Ising}} }=\frac{1}{2}
\label{eq:ratio2} \end{equation}
for the series $A_{0\ldots4}$, and by
\begin{equation}
\frac{ K^{2}_{\text{Ising}} }{ K^{1}_{\text{Ising}} }=\frac{1}{4}.
\label{eq:ratio4} \end{equation}
for the series $B_{0\ldots4}$.
Though we defined here ten additional points we actually performed
measurements only at the points $A_{2}$ and $B_{2}$.
 The points
$A_{2}$, $B_{2}$, $C_{2}$ represent three RBAT ( or AAT ) models with
coupling ratios $\frac{\Lambda^{1}}{ K^{1} } \approx \frac{1}{2}$ and
$\frac{ K^{2} }{ K^{1} } \approx \frac{1}{2},\;\frac{1}{4},\;\frac{1}{10}$
respectively.

\subsection{ MC Simulations}   \label{sec: MC}
We calculated the energy\cite{energy} per site:
 \begin{equation} E=[\av{E}]=-\frac{1}{L^{2}}[\av{
\sum_{<i,j>}\{K_{i,j}(\sigma_{i}\sigma_{j}+\tau_{i}\tau_{j})+
\Lambda_{i,j}\sigma_{i}\tau_{i}\sigma_{j}\tau_{j}\} } ]
    \label{eq:E}\end{equation}
where $L$ is the linear lattice size;
the angular brackets denote the usual thermal MC average whereas
the square brackets denote the quenched average over random-bond
configurations .
 The specific heat per site follows from the energy fluctuations
\begin{equation}
C=\frac{C_{h}}{k_{B}}=L^{2}[\av{E^{2}}-\av{E}^{2}] .
    \label{eq:Cv}\end{equation}
We define
\begin{equation}\begin{array}{lll}
m_{\sigma}= \frac{1}{L^{2}}\sum_{i}\sigma_{i} \;\;\; &
m_{\tau}= \frac{1}{L^{2}}\sum_{i}\tau_{i} \;\;\; &
m_{\sigma\tau}= \frac{1}{L^{2}}\sum_{i}\sigma_{i}\tau \;\;\; \\
\end{array} \;.\label{eq:m} \end{equation}
To compute the magnetization per site, $M$, we used
\begin{equation}
M=\frac{1}{2}[\av{ |m_{\sigma}|+|m_{\tau}| }] \,,
    \label{eq:M}\end{equation}
taking advantage of the symmetry between the $\sigma$ and $\tau$ spins to
increase accuracy.
Another order parameter of the transition plane we are considering is the
polarization\cite{exact}, $P$, defined as \begin{equation}
P=[\av{ |m_{\sigma\tau}| }] \,,
    \label{eq:P}\end{equation}
which measures the magnetization of the $\sigma\tau$ spins.
We calculated the magnetic susceptibility of the $\sigma$ spins
using\cite{energy}
  \begin{equation}
\chi=\frac{L^{2}}{2}[\av{ m_{\sigma}^{2}+ m_{\tau}^{2} } ]
  \label{eq:Xi}  \end{equation}
and also measured $c$, the size of the cluster flipped at each step, and
calculated $\av{ c }$. There is a connection between the size
of the clusters and the susceptibility \cite{Wolff:estimator} which
is common for all algorithms that generate non-interacting clusters of spins
(if all spins in a cluster have the same value). For the Wolff single
cluster procedure it has the simple form \cite{Wol:1C}
\begin{equation}
\chi=\av{c}   .
 \label{eq:improved estimator}  \end{equation}
Lastly we measured the susceptibility of the $\sigma\tau$ spins, $\chi_{p}$:
  \begin{equation}
\chi_{p}=L^{2}[\av{ m_{\sigma\tau}^{2} } ]
  \label{eq:XiP}  \end{equation}

 For each sample we estimated the effective autocorrelation
times\cite{S:Mon} of the energy,  $\tau_{E}^{\text{eff}}$ and the
magnetization, $\tau_{M}^{\text{eff}}$ using the method outlined in ref.
\onlinecite{Wol:1C}.
%\begin{equation}  \tau_{E,M}\approx
%\tau_{E,M}^{\text{eff}}\;, %\label{eq:tau} \end{equation}
%where $\tau_{E,M}^{\text{eff}}$\cite{eff} is estimated from the
%fluctuations of $E,M$.
The length of an individual run was often too short
for a precise estimate, but averaging over all
samples gave
a rough estimate  of $[\tau_{E,M}^{\text{eff}}]$. These estimates of
$\tau^{\text{eff}}$
allowed us to verify that all samples were equilibrated well enough before
we started collecting data. Typically we discarded at least $50
\tau_{E}^{\text{eff}}$ spin configurations.

The fluctuations in the random bond configurations introduce, in addition to
the usual MC error, another source of statistical error. Thus
the total error for the specific heat, $\delta C$, for example, is given by
\begin{equation}
(\delta C)^{2} =
\frac{\sigma^{2}_{n}}{n}+ \frac{\sigma^{2}_{T}} {n T_{MC}/\tau}
 \;, \;\;\;\;\;\;\;\;\;\;n,T_{MC}\;\;\mbox{  large.}
\label{eq:P toterr}\end{equation}
where $n$ is the number of random bond samples, $T_{MC}$ is the length of
the MC runs and $\tau$ is the autocorrelation time of the algorithm ( which
we can estimate by $\tau^{\text{eff}}$). $\sigma^{2}_{T}$ is the variance
of $C$ within a given sample and is due to thermal fluctuations, while
$\sigma^{2}_{n}$ is the variance of the exact $\av{C}$ within the ensemble
of random bond samples.
Thus both the thermal fluctuations and the sample to sample fluctuations
contribute to the error.

  We found that both terms in (\ref{eq:P toterr}) imply that simulations of
the AT models and the Potts model are much
more expensive in computer time than simulations of the Ising model.
 First there is an increase in $\tau$ in consistence with our
findings in the pure models\cite{SE}. E.g. at $L=256$ we find for
the random Ising model $\tau_{E}^{\text{eff}}\approx 3.1(1)$ compared with
$\tau_{E}^{\text{eff}}\approx 11.3(3)$ for the random four-state Potts
model.
  Second we found that for all thermodynamic quantities $P$ the relative
variance $\sigma^{2}_{n}/P^{2}$ was larger for the random four-state Potts
model than for the random Ising model. This effect was strongest for the
specific heat, e.g. for $L=256$ $\sigma^{2}_{n,C}/C^{2}$ of the random
four-state Potts model was approximately an order of magnitude larger
than that of the random Ising model.
 In accordance with these findings, the number of samples simulated for the
Ising model varied between $n= 1000$ for $L=4$ and $n=150$ for $L=256$,
while for the four-state Potts model it varied between $n= 2000$ for $L=4$
and $n=330$ for $L=256$.

\section{ Monte Carlo Results}     \label{sec:results}

\subsection{Specific Heat}
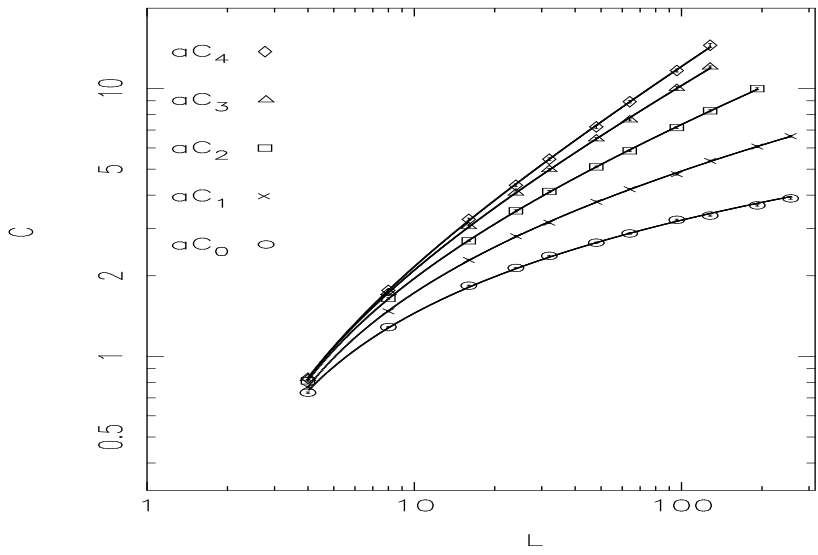
\begin{figure}[h]
\caption{ Specific heat, $C$,     as a function of $\log L$ for five
critical
models, $C_{0..4}$, of the AAT. Note that $C_{0}$ is an anisotropic
decoupled Ising model and $C_{4}$ is an anisotropic four-state Potts
model. The curves are fits to
( \protect\ref{eq: C(L)ani}) , yielding estimates for $\frac{ \alpha}{\nu}$
%of ( \protect\ref{eq:C fss})
 which are listed in table \ \protect\ref{tab: AAT1}
 . } \label{fig:CvAcFt1} \end{figure}

 We expect the critical specific heat of the AAT to follow the finite size
scaling\cite{Barber} form of (\ref{eq: C(t)ani}),
\begin{equation}
 C\approx a_{1}+b_{1} L^{\alpha/\nu}  \;.
\label{eq: C(L)ani} \end{equation}
 The exponent ratio $\frac{\alpha}{\nu}$ is predicted
analytically\cite{exact,Nienhuis} to
vary continuously along the critical line of the pure AT model between
$\frac{\alpha}{\nu}=0$ at the decoupled Ising critical point and
$\frac{\alpha}{\nu}=1$ at the four-state Potts critical point. One expects
the same variation to hold for the corresponding critical manifold of the
AAT at which the simulations were performed. In figure \ref{fig:CvAcFt1} we
plot the critical specific heat,
$C$ as a function of $\log L$ for the five critical AAT models $C_{0..4}$.
Fits
are made to the form (\ref{eq: C(L)ani}), and the estimated values of
$\alpha/\nu$ are listed in Table \ref{tab: AAT1}.
Since the model $C_{0}$
is an anisotropic Ising model, its specific heat does not follow the form
(\ref{eq: C(L)ani}), but rather $C\approx A+B\log L$. The anisotropic
four-state Potts model $C_{4}$ should have $\alpha/\nu=1$ with a logarithmic
correction\cite{Li:Sokal} to scaling $C\sim L/\log^{3/2} L$ . In the
range of $L$ considered, the logarithmic
correction results in an estimate for $\alpha/\nu$ which is too low.
  The results for the specific heat of the AAT models clearly demonstrate
the variation of $\alpha/\nu$ from its Ising value, $\alpha/\nu=0$ to its
four state Potts value.
As will be seen shortly this is quite different
from the results for the RBAT models.

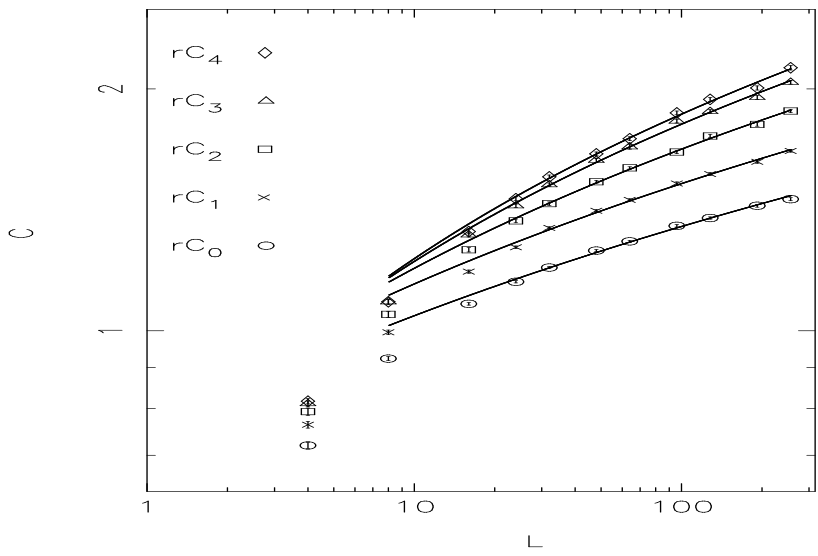
\begin{figure}[h]
\caption{ Specific heat, $C$,     as a function of $\log L$ for five
critical
models, $C_{0..4}$, of the RBAT. Note that $C_{0}$ is a random-bond
decoupled Ising model and $C_{4}$ is a random-bond four-state Potts
model. The curves are fits to
the form ( \protect\ref{eq: C(L)ani}) for latice sizes $L\geq 24$.
 } \label{fig:CvrcfC1} \end{figure}

 In order to analyze the specific heat results of the RBAT we first tried
 to fit the results to
(\ref{eq: C(L)ani}), but unlike the specific heat of the AAT, this was not
possible for the full lattice
size range $4\leq L\leq 256$. Thus in fig. \ref{fig:CvrcfC1} we show the
 specific
heat as a function of $\log L$ for the five critical RBAT models $C_{0..4}$
along with fitting curves to  (\ref{eq: C(L)ani}), where only data for
$L\geq24$ were used to fit.
The estimated exponents $\alpha/\nu$ obtained from these fits are 0 within
errors indicating a possible logarithmic behaviour. In addition, fig.
\ref{fig:CvrcfC1} indicates a crossover of the effective exponents (the
slope between every two successive data points) from higher to lower values.
Physically we expect that this is due to a crossover from the pure models'
exponents to some new random critical behaviour. We are thus led to the
following finite size scaling ansatz for the specific heat
 \begin{equation}
C=a_{0}+ b_{0} \ln [1+ c_{0}(L^{(\alpha/\nu)_{\text{p}}}-1 )]\;,
\label{eq:C fss cross} \end{equation}
 where $(\alpha/\nu)_{\text{p}}$ is the critical exponent ratio of the
corresponding pure model and $c_{0}=\frac{g}{ (\alpha/\nu)_{\text{p}} }$. An
important point to note is that for each random bond
AT model, following the discussion at the end of sec. \ref{sec:duality ran},
$(\alpha/\nu)_{\text{p}}$ is taken from our results
(listed in Table \ref{tab: AAT1}) for the corresponding anisotropic AT model.
$g$ is a measure of the randomness, and is related through the relation
\begin{equation}
c_{0}= (L_{c}^{(\alpha/\nu)_{\text{p}}}-1 )^{-1}
\;,  \label{eq:C Lc} \end{equation}
to a crossover length $L_{c}$,
 at which crossover from the pure model's power law behaviour to the random
logarithmic behaviour occurs. Thus for $ L \ll L_{c}$ our eq.
(\ref{eq:C fss cross}) reduces to (\ref{eq: C(L)ani}) while for $ L \gg
L_{c}$ and $ L^{(\alpha/\nu)_{\text{p}}}\gg 1$ a logarithmic behaviour is
attained,  \begin{equation}
C=a+ b \ln  L \;.
\label{eq:C fss log} \end{equation}

 Apart from crossing over to the correct pure result (\ref{eq: C(L)ani})
when $c_{0}\rightarrow 0$, eq. (\ref{eq:C fss cross})
was constructed so that in the Ising model limit, $(\alpha/\nu)_{\text{p}}
\rightarrow 0$, it becomes
 \begin{equation}
C=a+ b \ln (1+ g\ln L )\;.
\label{eq:C fss dotz} \end{equation}
This is the finite size scaling form of (\ref{eq: C(t)ising ran}) which was
shown by Wang et al.\cite{Wang1} to fit the random bond Ising model well.
This form will be examined below as another candidate for
describing the RBAT models.

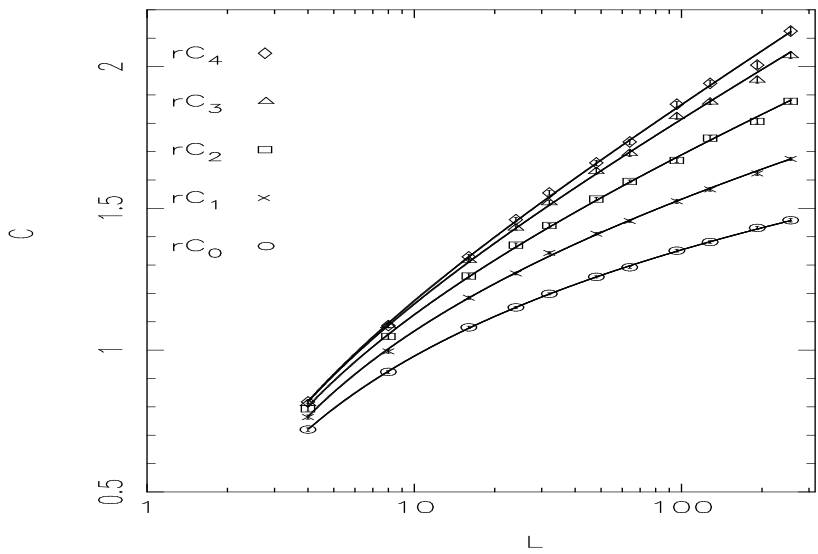
\begin{figure}[h]
\caption{ Specific heat, $C$,     as a function of $\log L$ for five
critical
models, $C_{0..4}$, of the RBAT. The curves are fits to
the form ( \protect\ref{eq:C fss cross}) , yielding estimates for the
coefficients of ( \protect\ref{eq:C fss cross})
 which are listed in table \ \protect\ref{tab:Cv fit}
 . } \label{fig:CvrCaat} \end{figure}

 In fig. \ref{fig:CvrCaat} the specific heat of the five critical RBAT
models $C_{0..4}$  are plotted again but with fits to (\ref{eq:C fss cross})
using the full  lattice size range $4\leq L\leq 256$, on a semilogarithmic
scale. The fitting parameters are listed in Table \ref{tab:Cv
fit} . For these points
( with large randomness, $\frac{ K^{2} }{ K^{1} } \approx
\frac{1}{10}$ ), the crossover lengths $L_{c}$ are found to be 1.

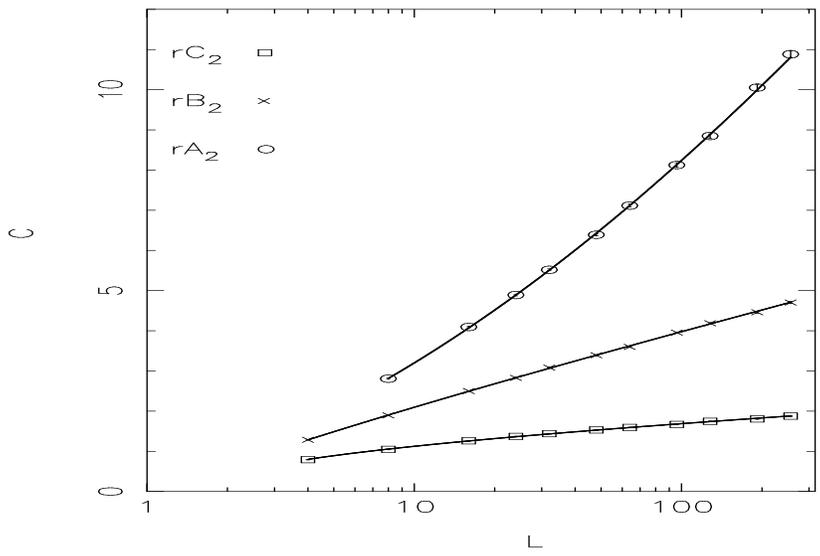
\begin{figure}[h]
\caption{ Specific heat, $C$,     as a function of $\log L$ for three
critical models,$A_{2},B_{2},C_{2}$, of the Random Bond AT .
The curves are
fits to the form ( \protect\ref{eq:C fss cross}) , yielding estimates for the
coefficients of ( \protect\ref{eq:C fss cross})
 which are listed in table \ \protect\ref{tab:Cv fit}
 . } \label{fig:Cvrdaat} \end{figure}

 In fig. \ref{fig:Cvrdaat} the specific heats of three critical RBAT
models $A_{2},B_{2},C_{2}$  are plotted with the fitting functions of
(\ref{eq:C fss cross}). $(\alpha/\nu)_{\text{p}}$ is taken from the
corresponding anisotropic models and the fitting parameters are listed in
Table \ref{tab:Cv fit} . $(\alpha/\nu)_{\text{p}}$ of the three models
$A_{2},B_{2},C_{2}$ is very similar ( .40, .37, .37 respectively) but they
differ in their amount of randomness,
$\frac{ K^{2} }{ K^{1} } \approx \frac{1}{2},\;\frac{1}{4},\;\frac{1}{10}$
respectively. The form (\ref{eq:C fss cross}) seems to describe all three
models adequately with the main change being in the crossover length
$L_{c}$ which decreases as randomness increases. The estimates for $L_{c}$
can be compared with results from random  bond Ising
 models\cite{rev selke,Wang1}  with the same coupling ratios. For  $\frac{
K^{2} }{ K^{1} } \approx \frac{1}{10},\;\frac{1}{4},\;\frac{1}{2}$ we obtain
$L_{c}=1,\;L_{c}=4.\pm.4,\;L_{c}=51\pm7$ compared with
$L_{c}=2\pm1,\;L_{c}=16\pm4$ and $L_{c}\sim 1000$ for the respective Ising
models. According to the Harris criterion one expects the crossover length
to scale roughly as
 \begin{equation}
 L_{c} \sim R^{-\frac{1}{\phi}}\;,
\label{eq:Lc Harris} \end{equation}
where $\phi=(\alpha/\nu)_{\text{p}}$ is the crossover exponent
 and $R$ ($R<1$, $R\sim \frac{ K^{1} }{ K^{2} }$)is a measure of the
randomness. For the RBAT
models considered here $(\alpha/\nu)_{\text{p}}\approx .37$, while for the
random bond Ising models randomness is marginal (
$(\alpha/\nu)_{\text{p}}=0$), so that the
much smaller crossover lengths that we find are consistent
with the Harris criterion.

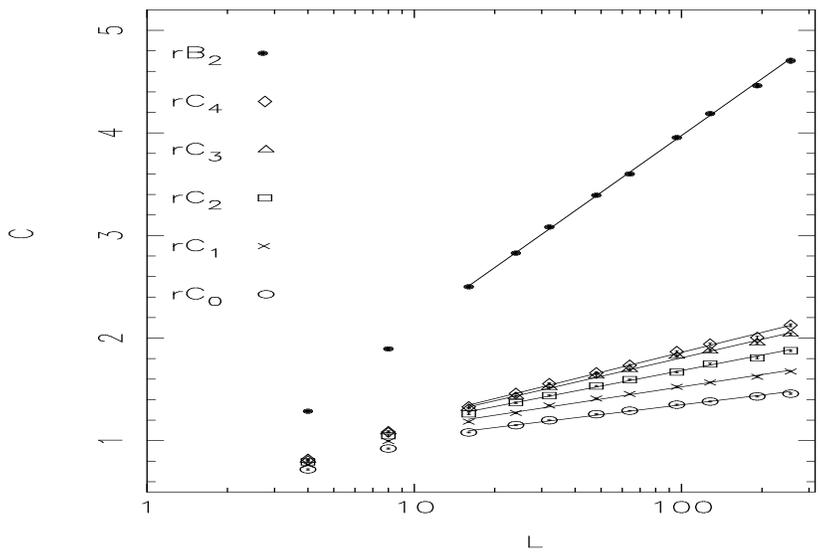
\begin{figure}[h]
\caption{ Specific heat, $C$,     as a function of $\log L$ for the RBAT
critical models,$B_{2}$ and $C_{0\dots 4}$ with fitting curves of
 ( \protect\ref{eq:C fss log}).
  } \label{fig: Cvrclg1} \end{figure}

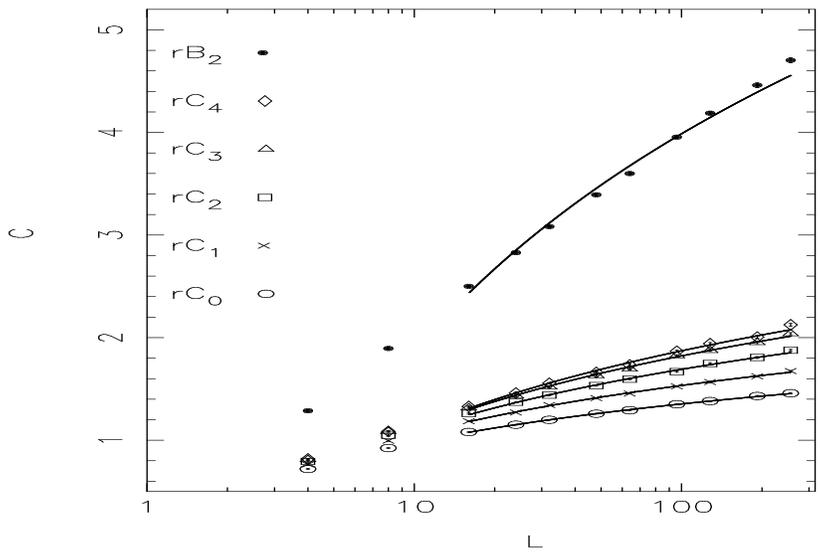
\begin{figure}[h]
\caption{ Specific heat, $C$,     as a function of $\log L$ for the RBAT
critical models,$B_{2}$ and $C_{0\dots 4}$ with fitting curves of
 ( \protect\ref{eq:C fss log log}).
  } \label{fig: CvrcLL1} \end{figure}

 Since there is no theoretical prediction for the critical behaviour of
the RBAT
model, we examined  possible scaling forms for the specific heat, other than
(\ref{eq:C fss cross}). A natural candidate is a double logarithmic
behaviour as in eq. (\ref{eq: C(t)ising ran}). This form was predicted
by DD for the disordered Baxter\cite{DD:bax} and Ising\cite{rev selke,Wang1}
models and confirmed for the latter model. In its asymptotic limit, the
finite size scaling form of (\ref{eq: C(t)ising ran}) is
 \begin{equation}
C=a+ b \ln \ln  L \;.
\label{eq:C fss log log} \end{equation}
We compare this form with (\ref{eq:C fss log}) as both these forms are
asymptotic and do not include crossover from the pure behaviour.
  In figure \ref{fig: Cvrclg1}  we plot again our
 specific heat results for all the models tested except for $A_{2}$ which
is not considered due to its large crossover length.
The results are fitted by (\ref{eq:C fss log}) in fig. \ref{fig: Cvrclg1}
and by (\ref{eq:C fss log log}) in fig. \ref{fig: CvrcLL1}, for $L \geq 16$.

As found by Wang
 et al., for the random bond Ising model $C_{0}$, agreement with (\ref{eq:C
fss log log}) is much better than with (\ref{eq:C fss log}), according to
our error analysis. For the model
$C_{1}$ too agreement with (\ref{eq:C fss log log}) is better but this can
be attributed to slow crossover behaviour from the small power law
$(\alpha/\nu)_{\text{p}}\approx .17$ ( the condition $
L^{(\alpha/\nu)_{\text{p}}}\gg 1$ is not fulfilled for such small
$(\alpha/\nu)_{\text{p}}$ and for the range of $L$ considered). For the
models $C_{2,3}$ the
quality of both fits is comparable while for the random bond four state Potts
$C_{4}$ the fit (\ref{eq:C fss log}) is better. For the model $B_{2}$
we have obtained larger statistics ( more than 900 samples for $L\leq 128$
 and 300 samples for $L=256$) and the result is
conclusively in favor of  (\ref{eq:C fss log}).

Last we considered the
possibility that the crossover from the power law behaviour to a logarithmic
behaviour is yet followed at even larger $L$ by a crossover to a
double logarithmic behaviour. This was done by fitting the specific heat data
 for $L\geq16$ according to (\ref{eq:C fss dotz}). Though the fits
were of good quality, the estimated crossover lengths $L_{i}=\exp(1/g)$ did
not make sense physically. For the $B_{2}$ model we found $L_{i}>10^{8}$
which actually implies a logarithmic behaviour for all accessible values
of $L$. This
should be compared against our estimate $L_{c}\approx4$ and the value
$L_{i}\approx 16$ for the corresponding Ising model (see above). For
the models $C_{0\ldots4}$ we find that $L_{i}$ increases monotonically from
the Ising model $C_{0}$ where $L_{i}\approx 2$  to the four state Potts model
$C_{4}$ where $L_{i}\approx200$. This is contrary to the expectation that
the crossover length should decrease due to the increase in the crossover
exponent $\phi=(\alpha/\nu)_{\text{p}}$, as implied from (\ref{eq:Lc
Harris}).

We conclude that our results are in
favor of a logarithmic divergence of the specific heat and not a double
logarithmic one for the RBAT model, excluding the random bond Ising model.
The finite size scaling form (\ref{eq:C fss cross}) gives the best agreement
with our results for all the models, including those with weak randomness,
and for the full lattice size range $4\leq L\leq 256$. Thus we predict for
$C(t)$ the scaling form:
 \begin{equation}
C\sim \ln [1+ b(t^{-\alpha_{\text{p}}}-1 )]\;,
\label{eq:C scaling cross} \end{equation}
  A possibility which we tend to rule out is that both
types of critical behaviors ( that is (\ref{eq:C scaling cross}) and
(\ref{eq: C(t)ising ran}) ) occur  and that
there is some  bifurcation plane which separates the two types of
behaviors.
%%%%%%%%%%%%%%%%%%%%%%%%%%%%%%%%%%%%%%%%%%%%%%%%%%%%%%%%%%%%%%%%%%%%%%%%%

\subsection{ Susceptibility and  Magnetization}
\subsubsection{ Susceptibility}
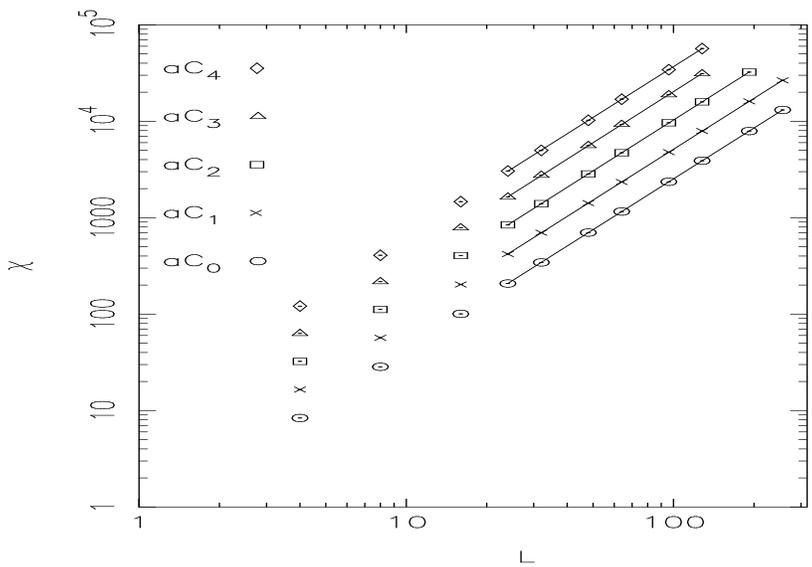
\begin{figure}[h]
\caption{ Susceptibility as a function of $\log L$ for five critical models,
$C_{0..4}$, of the AAT. For the sake of clarity $\chi$ for the model $C_{i}$
has been multiplied by $2^{i}$. The solid lines are linear fits to the form
( \protect\ref{eq:Xfss}) for $L\geq 24$, yielding estimates for critical
exponents
$\frac{\gamma}{\nu}$ which are listed in table \ \protect\ref{tab: AAT1} .
} \label{fig:XtAcall} \end{figure}

According to finite size scaling theory one expects that the critical
susceptibility diverges as
 \begin{equation}
\chi \sim L^{ \frac{\gamma}{\nu} }
  \label{eq:Xfss}  \end{equation}
for large enough $L$.
The exponents' ratio $\gamma /\nu$ is
predicted analytically\cite{exact,Nienhuis} to be
constant, $\gamma /\nu= \frac{7}{4}$, all along the critical line of the
pure AT model.
 This is expected also to hold for the corresponding critical
manifold of the AAT.
Figure \ref{fig:XtAcall} shows results for the critical susceptibility
as a function of $\log L$ for the five critical models, $C_{0..4}$, of the
AAT.
So that the points won't fall on top of each other, $\chi$ for the model
$C_{i}$ has been multiplied
by $2^{i}$. The solid lines are linear fits to the form (\ref{eq:Xfss}) for
$L\geq 24$, yielding estimates for critical exponents $\frac{\gamma}{\nu}$
which are listed in Table \ref{tab: AAT1}. All results seem to be consistent
with the analytic prediction, giving a combined best estimate of $\gamma
/\nu=1.7512(6)$. An effective exponent analysis (or a trend analysis) shows
that as $\log L$ increases, $(\frac{\gamma}{\nu})_{ \text{eff} } $
approaches the value $\frac{7}{4}$ from above.

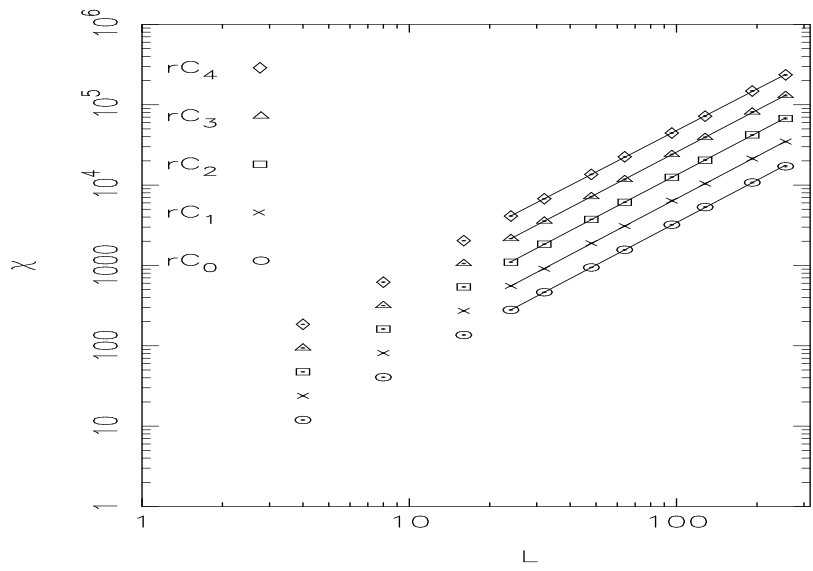
\begin{figure}[h]
\caption{ Susceptibility as a function of $\log L$ for five critical models,
$C_{0..4}$, of the RBAT. For the sake of clarity $\chi$ for the model
$C_{i}$
has been multiplied by $2^{i}$. The solid lines are linear fits to the form
( \protect\ref{eq:Xfss} ) for $L\geq 24$, yielding estimates for critical
exponents
$\frac{\gamma}{\nu}$ which are listed in table \ \protect\ref{tab: Rand1} .
} \label{fig:XtRcall} \end{figure}

 The finite size scaling form (\ref{eq:Xfss}) and the value of
$\frac{\gamma}{\nu}=\frac{7}{4}$ have been predicted\cite{Shal,Shan,Lud}
for the disordered decoupled Ising model and confirmed in MC
simulations\cite{rev selke,Wang1}.
Figure \ref{fig:XtRcall} shows results for the critical susceptibility
as a function of $\log L$ for the five random critical models $C_{0..4}$,
of the
RBAT. The same analysis has been carried out as for the AAT and the estimated
critical exponents $\frac{\gamma}{\nu}$
 are listed in Table \ref{tab: Rand1}. The estimates for the
decoupled Ising  critical point $C_{0}$ and for the models $C_{1,2}$ are
consistent with
 $\frac{\gamma}{\nu}=\frac{7}{4}$. On the other hand for the model $C_{3}$
 and the random-bond four-state Potts model $C_{4}$ we obtain
$\frac{\gamma}{\nu}=1.736(3)$
and $\frac{\gamma}{\nu}=1.714(5)$ respectively. We also tried fitting the
results with the form
  \begin{equation}
\chi = A_{\chi} L^{ \frac{7}{4} } (\ln L)^{-p_{\chi}} \;,
  \label{eq:Xlogfss}  \end{equation}
for $L\geq 24$. The obtained estimate for $p_{\chi}$ is listed in
Table \ref{tab: Rand1}. for the random-bond four state Potts we find
$p_{\chi}=.148(21)$ while for the models $C_{0-2}$ the error in $p_{\chi}$
is at least of the same order as the estimate for $p_{\chi}$ itself.

%%%%%%%%%%%%%%%%%%%%%%%%%%%%%%%%%%%%%%%%%%%%%%%%%%%%%%%%%%%%%%%%%%%%%%%%%%

\subsubsection{ Magnetization}

The critical magnetization of the five anisotropic models is
expected\cite{exact,Nienhuis} to follow the form
 \begin{equation}
M \sim L^{- \frac{\beta}{\nu} }\;,
  \label{eq:Mfss}  \end{equation}
  with $\frac{\beta}{\nu}= \frac{1}{8}$.
Applying straightforward finite size scaling analysis, as described
above for the susceptibility, to the magnetization results, we obtained
 estimates for $\frac{\beta}{\nu}$ which are listed in Table \ref{tab:
AAT1}.
 The results are slightly below the expected value, but
an effective exponent analysis (or a trend analysis) shows
that as $\log L$ increases $(\frac{\beta}{\nu})_{ \text{eff} } $ approaches
the value $\frac{1}{8}$ from below.

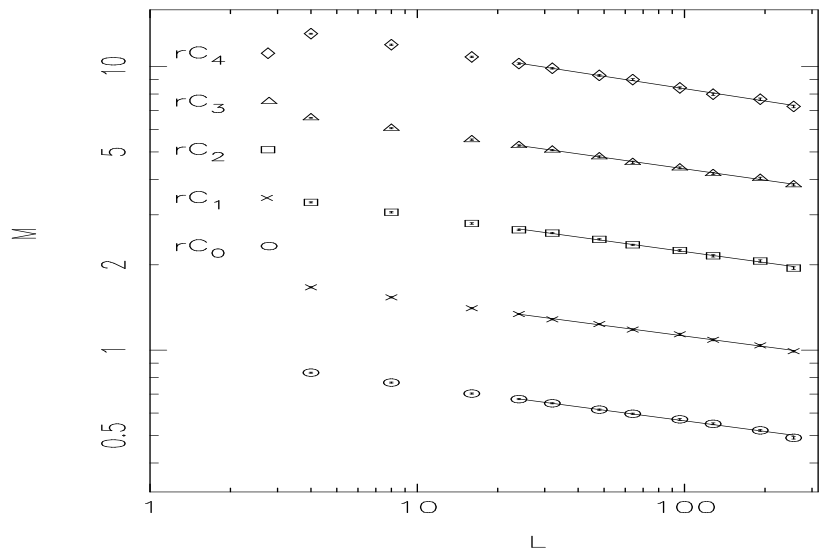
\begin{figure}[h]
\caption{ Magnetization as a function of $\log L$ for five critical models,
$C_{0..4}$, of the RBAT. For the sake of clarity $M$ for the model
$C_{i}$
has been multiplied by $2^{i}$. The solid lines are linear fits to the form
( \protect\ref{eq:Mfss} ) for $L\geq 24$, yielding estimates for critical
exponents
$\frac{\beta}{\nu}$ which are listed in table \ \protect\ref{tab: Rand1} .
} \label{fig:M1Rcall} \end{figure}

 The finite size scaling form (\ref{eq:Mfss}) and the value of
$\frac{\beta}{\nu}=\frac{1}{8}$ have been predicted\cite{Shal,Shan,Lud}
for the disordered decoupled Ising model and confirmed in MC
simulations\cite{rev selke,Wang1}.
Figure \ref{fig:M1Rcall} shows results for the critical magnetization
as a function of $\log L$ for the five critical models, $C_{0..4}$, of the
RBAT. A fit to the form (\ref{eq:Mfss}) yields the estimated
critical exponents $\frac{\beta}{\nu}$
 listed in Table \ref{tab: Rand1}. The estimates for the
decoupled Ising  critical point $C_{0}$ and for the models $C_{1,2}$ are
consistent with
 $\frac{\beta}{\nu}=\frac{1}{8}$. On the other hand for the model $C_{3}$
 and the random-bond four-state Potts model $C_{4}$ we obtain
$\frac{\beta}{\nu}=.133(2)$
and $\frac{\beta}{\nu}=.145(3)$ respectively. We also tried fitting the
results with the form
  \begin{equation}
M = A_{m} L^{ - \frac{1}{8} } (\ln L)^{-p_{m}} \;,
  \label{eq:Mlogfss}  \end{equation}
for $L\geq 24$. The obtained estimate for $p_{m}$ is listed in
Table \ref{tab: Rand1}. For the random-bond four state Potts model we find
$p_{m}=.082(13)$ while for the models $C_{0-2}$ the error in $p_{1}$ is
at least of the same order as the estimate for $p_{m}$ itself.

%%%%%%%%%%%%%%%%%%%%%%%%%%%%%%%%%%%%%%%%%%%%%%%%%%%%%%%%%%%%%%%%%%%%%%%%

\subsubsection{ Polarization and $\chi_{p}$ }
Since the polarization is an order parameter of the transition, the critical
polarization, $P$, and $\chi_{p}$ should follow the forms
\begin{mathletters}  \label{eq:PXp1}
 \begin{equation}
P\sim L^{ \frac{-\beta_{p}}{\nu} }   \;,
  \label{eq:P1} \end{equation}
 \begin{equation}
\chi_{p}\sim L^{  \frac{\gamma_{p}}{\nu} }
  \label{eq:Xp1}\;.  \end{equation}
\end{mathletters}

The critical exponents $\nu,\beta_{p}$ and $\gamma_{p}$ are known
analytically\cite{exact,Nienhuis} all along the critical line of the
pure AT model.
For example the ratio $\frac{\beta_{p}}{\nu}$ varies continuously between
its value at the
decoupled Ising critical point, $\frac{\beta_{p}}{\nu}=\frac{1}{4}$, and its
value at the four
state Potts critical point, $\frac{\beta_{p}}{\nu}=\frac{1}{8}$. These two
extreme values can
be obtained, through their relations with $\frac{\beta}{\nu}$, for any (
pure, anisotropic or disordered) decoupled Ising or four-state Potts model.
 For the decoupled Ising models
$\Lambda^{1}=\Lambda^{2}=0$ (or $\Lambda=0$ for the pure model), the
$\sigma$ spins system and the $\tau$ spins system are decoupled, hence
 \begin{equation}
P= \av{\sigma\tau}= \av{\sigma} \av{\tau}=M^{2} \sim t^{2\beta} \;.
\label{eq:PIsing} \end{equation}
This implies $\beta_{p}=2\beta$ for
all the decoupled Ising models.
%As the results for the model
%$C_{0}$ (both for the AAT and the RBAT ) indicate
%$\frac{\beta}{\nu}= \frac{1}{8}$ for decoupled Ising models, one expects
%$\frac{\beta_{p}}{\nu}=\frac{1}{4}$ for these models.
 For the four state Potts models $K^{1}=\Lambda^{1}$ and $K^{2}=\Lambda^{2}$
 (or $K=\Lambda$ for the pure four state Potts) so that $\sigma$
spins, $\tau$ spins and $\sigma\tau$ spins are equivalent, and
 \begin{equation}
P= \av{\sigma\tau}= \av{\sigma}= \av{\tau}=M \;.
\label{eq:PPotts} \end{equation}
It follows that $\beta_{p}=\beta$.
%The same consideration leads to the
%conclusion that $\gamma_{p}=\gamma$ for the four state Potts models. The
%scaling relation,
% \begin{equation}
%2\frac{\beta_{p}}{\nu}+\frac{\gamma_{p}}{\nu}=d  \;,
%\label{eq:scaling1} \end{equation}
%which our results fully confirm, leads to
%$\frac{\gamma_{p}}{\nu}=\frac{3}{2}$ for the Ising models.
%Thus if we
%accept the results for the susceptibility and the magnetization as
%indicating that
%$\frac{\gamma}{\nu}=\frac{7}{4}$ and
%$\frac{\beta}{\nu}=\frac{1}{8}$ for the critical plane of the RBAT then
Thus interpreting our results as $\frac{\beta}{\nu}=\frac{1}{8}$ for the
critical plane of the RBAT implies
$\frac{\beta_{p}}{\nu}=2\frac{\beta}{\nu}=\frac{1}{4}$ for the random-bond
decoupled Ising model, and
$\frac{\beta_{p}}{\nu}=\frac{\beta}{\nu}=\frac{1}{8}$ for the random-bond
four-state Potts model. The question still remains whether the variation
between these two values along the critical plane of the RBAT will be
continuous, as is the case for the pure model, or whether it will become
discontinuous under the influence of disorder.
The results presented below are in favor of a continuous variation.
The values of $\frac{\gamma_{p}}{\nu}$ can be obtained through the scaling
relation,
 \begin{equation}
2\frac{\beta_{p}}{\nu}+\frac{\gamma_{p}}{\nu}=d  \;,
\label{eq:scaling1} \end{equation}
which our results fully confirm. This leads to
$\frac{\gamma_{p}}{\nu}=\frac{3}{2}$ for the Ising models and
$\frac{\gamma_{p}}{\nu}=\frac{7}{4}$ for the four-state Potts models.

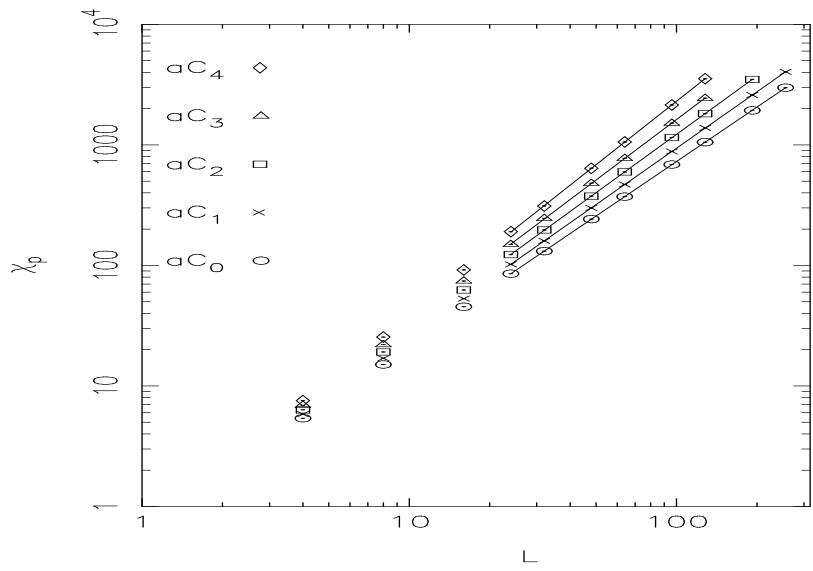
\begin{figure}[h]
\caption{ $\chi_{p}$     as a function of $\log L$ for five critical models,
$C_{0..4}$, of the AAT. The solid lines are linear fits to the form
( \protect\ref{eq:Xp1}) for $L\geq 24$, yielding estimates for critical
exponents
$\frac{\gamma_{p}}{\nu}$ which are listed in table \ \protect\ref{tab: AAT1}
. } \label{fig:XPAcall} \end{figure}

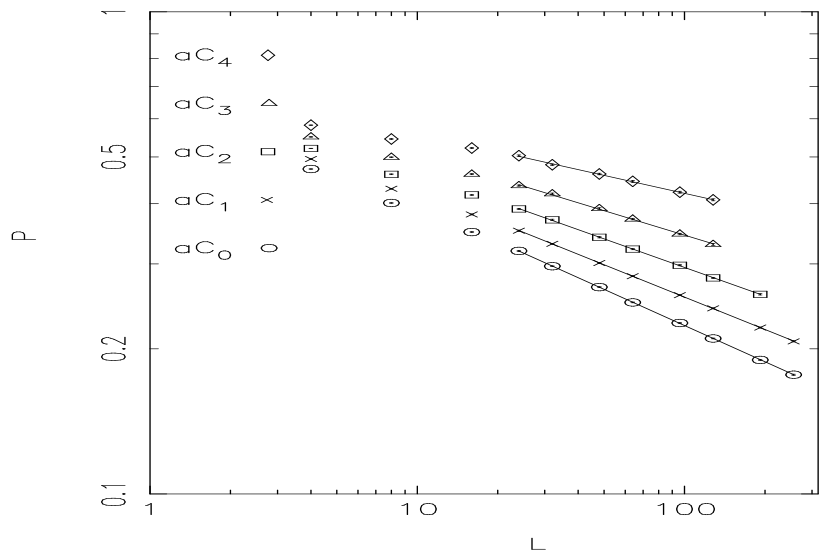
\begin{figure}[h]
\caption{ Polarization $P$  as a function of $\log L$ for five critical
models, $C_{0..4}$, of the AAT. The solid lines are linear fits to the form
( \protect\ref{eq:P1}) for $L\geq 24$, yielding estimates for critical
exponents
$\frac{\beta_{p}}{\nu}$ which are listed in table \ \protect\ref{tab: AAT1}
. } \label{fig:P3Acall} \end{figure}

Figures \ref{fig:XPAcall} and \ref{fig:P3Acall} show  the
critical $\chi_{p}$ and polarization, $P$, respectively,
as a function of $\log L$ for the five critical models, $C_{0..4}$, of the
AAT.
The solid lines are linear fits to the forms (\ref{eq:PXp1}) for
$L\geq 24$, yielding estimates for critical exponents
$\frac{\gamma_{p}}{\nu}$ and $\frac{\beta_{p}}{\nu}$
which are listed in Table \ref{tab: AAT1}. As expected we find that
$\frac{\gamma_{p}}{\nu}$ varies between $\frac{\gamma_{p}}{\nu}=1.501(1)$
for the decoupled Ising model and $\frac{\gamma_{p}}{\nu}=1.750(4)\approx
\frac{\gamma}{\nu}$ for the four state Potts model. $\frac{\beta_{p}}{\nu}$
varies accordingly fulfilling the scaling relation ( \ref{eq:scaling1} ).

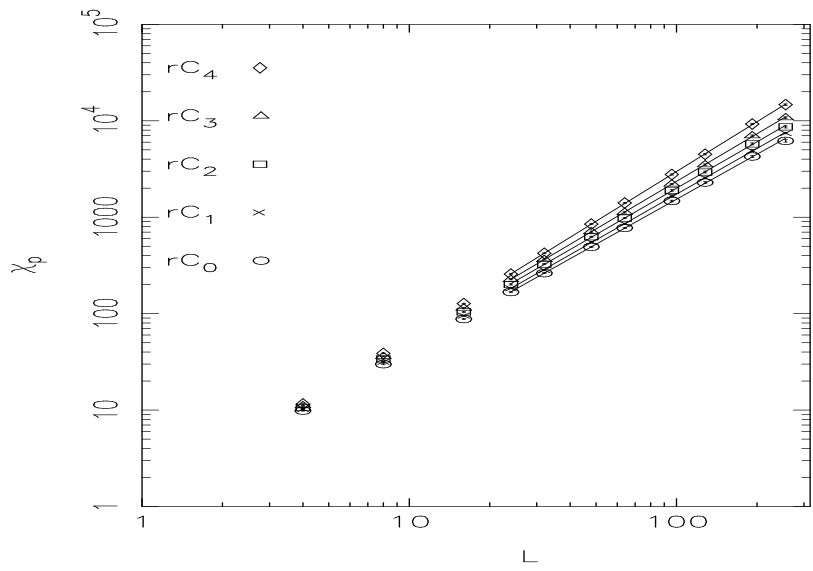
\begin{figure}[h]
\caption{ $\chi_{p}$     as a function of $\log L$ for five critical models,
$C_{0..4}$, of the RBAT. The solid lines are linear fits to the form
( \protect\ref{eq:Xp1}) for $L\geq 24$, yielding estimates for critical
exponents
$\frac{\gamma_{p}}{\nu}$ which are listed in table \ \protect\ref{tab:
Rand1} . } \label{fig:XPRcall} \end{figure}

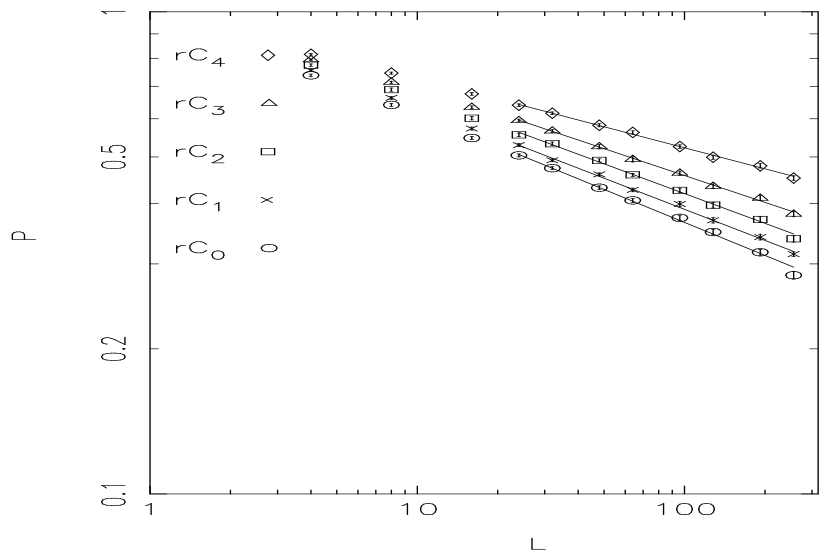
\begin{figure}[h]
\caption{ Polarization $P$  as a function of $\log L$ for five critical
models, $C_{0..4}$, of the RBAT. The solid lines are linear fits to the
form ( \protect\ref{eq:P1}) for $L\geq 24$, yielding estimates for critical
exponents
$\frac{\beta_{p}}{\nu}$ which are listed in table \ \protect\ref{tab: Rand1}
. } \label{fig:P3Rcall} \end{figure}

Figures \ref{fig:XPRcall} and \ref{fig:P3Rcall} show the
critical $\chi_{p}$ and polarization, $P$, respectively,
as a function of $\log L$ for the five critical models, $C_{0..4}$, of the
RBAT. The same analysis has been carried out as for the AAT and the estimated
critical exponents $\frac{\gamma_{p}}{\nu}$ and $\frac{\beta_{p}}{\nu}$
 are listed in Table \ref{tab: Rand1}. The results for the random-bond
four-state Potts, $C_{4}$, agree with $P=M$ and $\chi_{p}=\chi$ to a high
precision  and thus need no further explanation.

There is a clear mismatch between the estimated
values of $\frac{\gamma_{p}}{\nu}$
and $\frac{\beta_{p}}{\nu}$ and their expected values for the random-bond
Ising model $C_{0}$.
Even the left side of
the relation (\ref{eq:PIsing}), $P=M^{2}$, which implied
$\frac{\beta_{p}}{\nu}=2\frac{\beta}{\nu}$, is far from being met, but rather
we find $P>M^{2}$ both
for the anisotropic and for the random-bond Ising models. This is probably
connected to the fact that we are measuring, as is usual in a MC
calculation, $\av{|M|}$ and $\av{|P|}$ which
are larger than $\lim_{ h\rightarrow 0} \av{M}$ and $\lim_{ h_{p}\rightarrow
0} \av{P}$ respectively.
Both in the random-bond and in the anisotropic Ising model the relative
variance of $P$ is larger than the relative variance of $M$. This makes
$\av{|P|}$ a worse estimator of $\lim_{ h_{p}\rightarrow0} \av{P}$ than
$\av{|M|}$ is of $\lim_{ h\rightarrow 0} \av{M}$.
%The results for the $P$ for the random-bond Ising show a relative
%variance (due to sample to sample fluctuations ) twice as large as the
%relative variance of $M$. We can try to interpret the sample to sample
%fluctuations of observables as being due to fluctuations in $T_{c}$ of
%samples. If these fluctuations are large then tagadadam
The discrepancy is a finite size effect\cite{BH:book} and should decay as
$L$ increases.
%The discrepancy could also result from slow convergence to the asymptotic
%behaviour, as can be noticed by careful inspection of figures
%\ref{fig:XPRcall} and \ref{fig:P3Rcall}.
Indeed if
we estimate $\frac{\beta_{p}}{\nu}$ from lattices of size $L\geq 96$ we obtain
$\frac{\beta_{p}}{\nu}=.266(22)$ which is consistent with the exact value
$\frac{\beta_{p}}{\nu}=\frac{1}{4}$. An estimate of $\frac{\gamma_{p}}{\nu}$
from lattices of size $L\geq 64$ yields $\frac{\gamma_{p}}{\nu}=1.524(23)$
which is consistent with the exact value
$\frac{\gamma_{p}}{\nu}=\frac{3}{2}$.

The results for $\chi_{p}$ and $P$  seem to favor a continuous
variation of $\frac{\beta_{p}}{\nu}$ and $\frac{\gamma_{p}}{\nu}$ for
 the RBAT models as is the case for the AAT models. As explained above, for
symmetry reasons one expects that if
$\beta_{\text{Ising}}=\beta_{\text{Potts}}$ then $\beta_{p,\text{Ising}}\neq
\beta_{p,\text{Potts}}$. This is consistent with the variation of exponents
we find. None the less since we found the specific heat of all models to
diverge logarithmically one could expect the RBAT models to be more
universal. A possible scenario to expect could be
$\beta_{p}=\beta_{p,\text{Potts}}$ for all RBAT models but the Ising model.
Then assuming $(\beta/\nu)_{\text{Ising}}=1/8$ with (\ref{eq:scaling1})
would imply $\frac{\gamma_{p}}{\nu}=3/2$ for the Ising model and
$\frac{\gamma_{p}}{\nu}=7/4$ for all the other RBAT models. Such a
behaviour would of course be masqueraded by crossover, but none the less we
do not find evidence for such a behavior in our results. Further
investigation of this problem is clearly needed.
\section{ Summary and Discussion }    \label{sum}

 We've examined the effect of bond-disorder on the line of critical points
of the AT model. A
duality transformation was used in order to locate a critical
manifold of the random bond AT model which corresponds to the line of
critical points of the pure model. An anisotropic AT model was used as
a convenient reference pure model. Our results consist of the finite size
dependence of $C,M,P,\chi,\chi_{p}$ at criticality of several random bond
and anisotropic AT models. The models $C_{0\ldots 4}$ interpolate between
Ising and four-state Potts models with large disorder, thus also monitoring
the effect of changing the crossover exponent $(\alpha/\nu)_{\text{p}}$. The
models
$A_{2},B_{2},C_{2}$ serve to monitor the effect of changing the amount of
disorder while keeping the crossover exponent approximately the same (
$.37\leq(\alpha/\nu)_{\text{p}}\leq.40$ ).

   The results of the critical specific heat of all the RBAT models agree
very
well with the crossover behaviour (\ref{eq:C fss cross}), according to which
for $L<L_{c}$, $C$ diverges with the same exponent
$(\alpha/\nu)_{\text{p}}$ as
the corresponding AAT model, while for $L>L_{c}$, $C$ grows as $\log L$.
Our estimates for the crossover lengths of the models $A_{2},B_{2}$ are
much smaller than estimates of Wang et al. for random bond Ising models
with the same amount of disorder. This is consistent with the Harris
criterion and the crossover
exponent $(\alpha/\nu)_{\text{p}}\approx.4$ of the former models and the
crossover exponent $(\alpha/\nu)_{\text{p}}=0$ for the latter models.
 For the random bond Ising model we find a double logarithmic divergence
(\ref{eq:C fss dotz}), or (\ref{eq:C fss log log}) asymptotically, as was
found by Wang et al.
We view the adequacy of (\ref{eq:C fss cross}) for all the RBAT models (
with different $(\alpha/\nu)_{\text{p}}$ and different amounts of
randomness) for the full lattice range $\leq L \leq 256$, together with the
consistency of the
crossover lengths with physical understanding (Harris criterion), as a
strong
indication that the RBAT models ( including the four state Potts model )
exhibit a a logarithmic divergence of $C$.
We tend to rule out the possibility of a double logarithmic divergence of
$C$ as predicted by DD\cite{DD:bax} for the Baxter model, for reasons
explained in sec. \ref{sec:results}.

Although $\beta/\nu$ and $\gamma/\nu$ do not vary for the pure model, our
results indicate a possibility that $\beta/\nu$ and $\gamma/\nu$ do vary for
the random models. None the less this very small variation might be an
artifact of a correction to scaling, e.g. of a logarithmic type.
The exponents connected with the polarization $P$ and the susceptibility
$\chi_{p}$, $\beta_{p}/\nu$ and $\gamma_{p}/\nu$ which vary continuously for
the pure AT model seem to do so also for the RBAT model. But on the basis of
universality the possibility that these exponents are actually the same (with
some non-universal
corrections to scaling and a crossover effect ) for all RBAT models
excluding the Ising
model, seems to be a plausible scenario in need of further investigation.

Very recently an experimental study of a two dimensional four-state Potts
system with quenched impurities was done\cite{experimental} and the values
of $\nu=1.03(8)$,
$\beta=.135(1)$, and $\gamma=1.68(15)$ were obtained. These are consistent
with our
findings for the four-state Potts system $C_{4}$ but with a value of $\beta$
which is smaller than ours.

% Var:
%Practically this means that if
%one wants to make measurements in a critical
%disordered system of a quantity which diverges as $l^{\omega}$ to some
%given relative accuracy, the number of samples needed will decrease with
%$L$ only if $\alpha/\nu<0$.

\acknowledgments
SW would like to thank H.J. Heermann and the Many Particle Group of the HLRZ
J\"{u}lich for their warm hospitality and the generous allocation of
computer time on the Intel IPSC/860 and the Paragon.
This research has been supported by the US-Israel Bi-national Science
Foundation (BSF), and the Germany-Israel Science Foundation (GIF).

\appendix
\section*{The Duality Transformation of the AT Model} \label{app:embed}
We list here the duality transformation $\vec{D}[K,\Lambda]$ of the AT model
and a few of its properties. For convenience we use instead of
$(K,\Lambda)$ different variables for the coupling constants,
$(X,Z)$;
\begin{equation}   \begin{array}{lll}
 Z=\exp^{-4K}   & X=\exp^{-2(K+\Lambda)} & \; \vec{X}=(X,Z).
\end{array} \label{eq:def xzapp} \end{equation}
 Under a duality transformation
$\vec{D}$,
 a bond of strength $\vec{X}=(X,Z)$ transforms into a bond on the dual
lattice of
strength  $\vec{\tilde{X}}=(\tilde{X},\tilde{Z})=\vec{D}(\vec{X})$ given by
\cite{Wu:Wang,Domany:Riedel}
\begin{equation} \begin{array}{l}
\tilde{X}=(1-Z)/\Delta           \\
\tilde{Z}=(1-2X+Z)/\Delta        \\
\Delta=1+2X+Z                    \end{array} . \label{eq:duality}
\end{equation}
We list several properties of the transformation. It is self-inverse, i.e.,
$\vec{D}[\vec{D}(\vec{X})]=\vec{X}$. It maps the zero temperature point (
$\vec{X}=(0,0)$ ) onto $T=\infty$ ( $\vec{X}=(1,1)$ ) and vice versa.
The Ising subspace $Z=X^{2}$ and the 4-state
Potts subspace $Z=X$ are invariant under $\vec{D}$.
The subspace $\Sigma$, defined by $X\geq Z\geq X^{2}$ ( or $K\geq
\Lambda\geq 0$ ), is also invariant under $\vec{D}$.
  The line $Z=1-2X$ is a
self dual line and its intersection with $\Sigma$ is the line of critical
points of the AT model. It is also
true that if $Z\leq 1-2X$ then $\tilde{Z} \geq1-2\tilde{X}$ and vice versa.

% now the references. delete or change fake bibitem. delete next three
%   lines and directly read in your .bbl file if you use bibtex.
%\begin{references}
%\bibitem{tag} Fake bibitem.
%\end{references}

% \begin{figure}
% \caption{}
% \label{}
% \end{figure}

%begin{figure}[h]
%caption{ Specific heat, $C$,     as a function of $\log L$ for five
%ritical
%models, $C_{0..4}$, of the RBAT. Note that $C_{0}$ is a random-bond
%decoupled Ising model and $C_{4}$ is a random-bond four-state Potts
%model. The curves are fits to
%the form ( \protect\ref{eq:C fss}) , yielding estimates for the
%coefficients of ( \protect\ref{eq:C fss})
% which are listed in table \ \protect\ref{tab:Cv fit}
% . } \label{fig:CvDotz RC} \end{figure}

%%%%%%%%%%%%%%%%%%%%%%%%%%%%%%%%%%%%%%%%%%%%%%%%%%%%%%%%%%%%%%%%%%%
 \begin {table}
\caption{  Critical exponents ratios and fitting parameters from the first
series of simulations, $C_{0..4}$, and the second series $A,B,C_{2}$ of the
Anisotropic AT model. The exponent ratios and parameters are defined in the
text.} \label{tab: AAT1}
\begin{tabular}{llllll}
       & $\frac{\alpha}{\nu}$ & $\frac{\gamma}{\nu}$ & $\frac{\beta}{\nu}$
&$\frac{\gamma_{p}}{\nu}$ &$\frac{\beta{p}}{\nu}$ \\ \hline
$C_{0}$& .0001(150)& 1.7511(7) & .1244(4) & 1.501(1)& .2494(6)\\
$C_{1}$& .171(5)   & 1.751(1)  & .1242(5) & 1.554(1)& .2231(5)\\
$C_{2}$& .375(5)   & 1.755(1)  & .1223(7) & 1.608(2)& .197(1) \\
$C_{3}$& .549(8)   & 1.759(4)  & .1199(27)& 1.667(5)& .169(3) \\
$C_{4}$& .630(8)   & 1.749(4)  & .1238(21)& 1.750(4)& .1237(26) \\
$B_{2}$& .371(5)   & 1.753(1)  & .1233(7) & 1.608(2)& .197(1)   \\
$A_{2}$& .40(1)    & 1.750(1)  & .1254(8) & 1.603(2)& .199(1)   \\
 \end{tabular}
\end{table}

 \begin {table}
\caption{  Fitting parameters for the fits in figures
\protect\ref{fig:CvrCaat} and \protect\ref{fig:Cvrdaat} according to
equations
( \protect\ref{eq:C fss cross} , \protect\ref{eq:C Lc}). Errors are given
in parentheses only when error is smaller than or of the same order as the
number itself. }
\label{tab:Cv fit}
\begin{tabular}{lllll}
       & $a_{0}$ & $b_{0}$ & $c_{0}$     & $L{c}$ \\ \hline
$C_{0}$& -.37(12)& .58(1)  & 5.2E4(1.5E4)&  1.     \\
$C_{1}$& -4.6    & .51(2)  & 1.5E6       &  1.     \\
$C_{2}$& -4.1    & .46(127)& 5.5E4       &  1.     \\
$C_{3}$& -3.9    & .43(4)  & 5.5E4       &  1.     \\
$C_{4}$& -4.1    & .42(1)  & 1.0E5       &  1.     \\
$B_{2}$& -.09(5) & 2.00(4) & 1.47(10)    &  4.0(4) \\
$A_{2}$& -.07(6) & 9.35(33)&  .26(2)     &  51(7)
 \end{tabular}
\end{table}

% \begin {table}
%\caption{  Results from the AT critical line. The errors in parentheses are
%only the statistical errors of the fit in our fitting interval. They do not
%include systematic errors stemming from finite size effects and corrections
%to scaling.}
%\label{tab:Cv fit dotz}
%\begin{tabular}{lllll}
%       & $c_{0}$ & $c_{1}$ & $b$     & $L{i}$ \\ \hline
%$C_{0}$& -.37(12)& .58(1)  & 3.9(1.2)&  1.29\\
%$C_{1}$& -.15(9) & .82(4)  & 1.5(3)  &  1.95 \\
%$C_{2}$& .06(4)  & 1.28(7) & .55(8)  &  6.16\\
%$C_{3}$& .13(3)  & 1.8(1)  & .33(4)  &  20.7\\
%$C_{4}$& .16(4)  & 2.32(25)& .24(4)  &  64.5\\
%$B_{2}$& -.00(2) & 19.(3)  & .050(9) &  4.8E08
%\end{tabular}
%\end{table}
%%%%%%%%%%%%%%%%%%%%%%%%%%%%%%%%%%%%%%%%%%%%%%%%%%%%%%%%%%%%%%%%%%%%%%

 \begin {table}
\caption{  Critical exponents ratios and fitting parameters from the first
series of simulations, $C_{0..4}$, and the second series $A,B,C_{2}$ of the
Random Bond AT. These exponents and parameters related to the scaling of the
order parameters are defined in the text.}
 \label{tab: Rand1}
\begin{tabular}{lllllll}
       & $\frac{\gamma}{\nu}$ & $p_{\chi}$ & $\frac{\beta}{\nu}$ & $p_{m}$
& $\frac{\gamma_{p}}{\nu}$ & $\frac{\beta_{p}}{\nu}$\\ \hline
$C_{0}$& 1.751(5)& -.007(20)&  .125(3)& -.003(12)& 1.549(9)& .227(5)\\
$C_{1}$& 1.751(6)& -.008(19)&  .124(3)& -.004(11)& 1.575(8)& .214(5)\\
$C_{2}$& 1.743(5)&  .026(22)&  .129(3)&  .015(14)& 1.597(9)& .205(5)\\
$C_{3}$& 1.736(3)&  .057(14)&  .133(2)&  .032(9) & 1.638(5)& .185(3)\\
$C_{4}$& 1.714(5)&  .148(21)&  .145(3)&  .082(13)& 1.714(5)& .145(3)\\
$B_{2}$& 1.738(4)& -.049(17)&  .132(3)&  .029(10)& 1.586(6)& .209(4)\\
$A_{2}$& 1.739(5)& -.042(22)&  .132(3)&  .027(13)& 1.590(8)& .208(5)
 \end{tabular}
\end{table}


\begin{thebibliography}{99}
\bibitem[*]{email1} email address: shai@elect1.weizmann.ac.il
\bibitem[\dagger]{email2} email address: fedomany@WEIZMANN.weizmann.ac.il
\bibitem{Harris} A.B. Harris {\em J.Phys. C},{\bf7}:1671,(1974).
\bibitem{Stinch} R.B. Stinchcombe , in {\em Phase Transitions and
CriticalPhenomena},edited by C. Domb and J.L. Lebowitz(Academic, New York,
1983), vol. 7.
\bibitem{Kinz:Dom} W.Kinzel and E. Domany {\em Phys. Rev. B},{\bf23}:3421,
(19801).
\bibitem{L and H} T.C. Lubensky and A.B. Harris {\em AIP Conf. Proc.},
{\bf24}:311,(1974).
% \bibitem{L} T.C. Lubensky {\em Phys. Rev. B},{\bf11}:3573,(1975).
 \bibitem{J and K} C. Jayaprakash and H.J. Katz {\em Phys. Rev. B},{\bf16}
 :3987,(1977).
 \bibitem{Dots:Ising} Vik S. Dotsenko and Vl S. Dotsenko, Adv. Phys.
{\bf32}, 129(1983).
 \bibitem{Shal} B.N. Shalaev, Sov. Phys. Solid State {\bf26}, 1811 (1984).
 \bibitem{Shan} R. Shankar, Phys Rev. Lett. {\bf58}, 2466 (1987); Phys Rev.
Lett. {\bf61}, 2390 (1988).
  \bibitem{Lud} A.W.W. Ludwig, Phys Rev. Lett. {\bf61}, 2388 (1988).
 \bibitem{rev selke} W. Selke , in Computer Simulations in Condensed
Matter Physics III , edited by D.P. Landau,K.K. Mon, and H.B. Sch\"{u}ttler
(Springer, Heidelberg, 1991).
 \bibitem{Andelman} D. Andelman and A. Nihat Berker {\em Phys. Rev.
B},{\bf29}:2630,(1984).
 \bibitem{Der Gar} B. Derrida and E. Gardner, {\em J. Phys A},{\bf17}, 3223
(1984).
\bibitem{Lud Q1} A.W.W. Ludwig, {\em Nuc. Phys. B}{\bf285}, 97 (1986).
 \bibitem{8potts} S. Chen, A.M. Ferrenberg and D.P. Landau {\em Phys. Rev.
Lett.},{\bf69}:1213,(1992).
 \bibitem{Berker} A.N. Berker {\em Physica A }{\bf194}, 72 (1993).
 \bibitem{Nov Lan} M.A. Novotny and D.P. Landau {\em Phys. Rev. B},{\bf24}
:1468,(1981).
  \bibitem{Baxter Wu} R.J. Baxter and F.Y. Wu {\em Phys. Rev.
Lett.},{\bf31}:1294,(1973).
% \bibitem {eff} E.K. Riedel and F.J. Wegner {\em Phys. Rev. B},{\bf9}:294,
%(1974).
\bibitem{AT:1} J. Ashkin and E. Teller, {\em Phys. Rev.}{\bf64,} 178 (1943).
\bibitem{Wu:Lin} F.Y. Wu and K.Y. Lin, {\em J.Phys. C}{\bf7,} L181 (1974).
\bibitem{Domany:Riedel} E. Domany and E.K. Riedel, {\em Phys. Rev. B}{\bf
19,} 5817 (1979).
\bibitem{AT3d} R.V. Ditzian, J.R. Banavar, G.S. Grest and L.P. Kadanoff,
{\em Phys. Rev. B}{\bf22,} 2542, (1980).
\bibitem{SE} S. Wiseman and E. Domany, {\em Phys. Rev E}{\bf48,} 4080
(1993).
\bibitem{exact} R.J. Baxter {\em Exactly Solved Models in Statistical
Mechanics}, Academic press (1982).
\bibitem{Nienhuis} B. Nienhuis  , in {\em Phase Transitions and Critical
Phenomena},
edited by C. Domb and J.L. Lebowitz(Academic, New York, 1987), vol. 11.
 \bibitem{Alcaraz} F.C. Alcaraz and C. Tsallis, J.Phys. A {\bf15}, 587
(1982).
\bibitem{Baxter} R.J. Baxter, Ann. Phys. (N.Y.) {\bf70}, 193 (1972).
\bibitem{DD:bax} Vik S. Dotsenko and Vl S. Dotsenko {\em J.Phys.
A}{\bf17}:L301,(1984).
\bibitem{MLS} D. Matthews-Morgan, D.P. Landau and R.H. Swendsen  {\em Phys.
Rev. Lett.},{\bf53}:679,(1984).
\bibitem{rev:clust} For a review see: A.D. Sokal, {\em Nuc. Phys. B (Proc.
Suppl.)},{\bf20,}55, (1991)
\bibitem{BH:book} K. Binder, D.W. Heermann, {\em Monte Carlo Simulations
 in Statistical Physics. } Springer-Verlag, Berlin, (1988).
 \bibitem{Heuer} H.O. Heuer, Phys. Rev. B {\bf42}, 6476 (1990).
 \bibitem{Wang1} J.S. Wang, W. Selke, VI.S. Dotsenko and V.B. Andreichenko
 {\em Physica A},{\bf164}:221,(1990).
\bibitem{Hen} M. Hennecke and U. Heyken, {\em J. Stat. Phys.}{\bf 72}, 829
(1993).
\bibitem{Wu:Wang} F.Y. Wu and Y.K. Wang, J. Math. Phys. {\bf17}, 439
(1976).
 \bibitem{Fisch} R. Fisch {\em J. Stat. Phys.},{\bf18}:111,(1978).
\bibitem{Wol:1C} U. Wolff, {\em Phys. Rev. Lett.}{\bf62,} 361 (1989).
% \bibitem{DK:Gen} D. Kandel and E. Domany, {\em Phys. Rev. B}{\bf43,} 8539
%(1991).
\bibitem{embed:1} S. Caracciolo, R.G. Edwards, A. Pelissetto, and A.D.
Sokal, {\em Nuc. Phys. B (Proc. Suppl.)}{\bf20,} 72 (1991).
\bibitem{SW:1} R.H. Swendsen and J.S. Wang, Phys. Rev.
Lett. {\bf58}, 86 (1987).
\bibitem{r=.5} V.B. Andreichenko, Vl.S. Dotsenko, W.Selke, and J.-S. Wang,
Nucl. Phys. B {\bf344}, 531 (1990).
\bibitem{energy} A factor of $k_{B}T$ has been omitted,
but is irrelevant since all measurements are at a single temperature.
\bibitem{Wolff:estimator} U. Wolff {\em Nucl. Phys. B},{\bf300}:501,(1988).
\bibitem{S:Mon} A.D. Sokal, in {\em Computer Simulation Studies in Condensed
Matter Pysics: Recent Developments, }ed. D.P. Landau, K.K. Mon and H.-B.
\bibitem{Barber} M.N. Barber, in {\em Phase Transitions and Critical
Phenomena}, edited by C. Domb and J.L. Lebowitz, vol. 8.
\bibitem{Li:Sokal} X.J. Li and A.D. Sokal, {\em Phys. Rev.
Lett.}{\bf63,} 827 (1989).
\bibitem{experimental} L. Schwenger, K. Budde, C. Voges, and H. Pfn\"{u}r,
{\em Phys. Rev. Lett.}{\bf73}, 296 (1994).
  \end{thebibliography}
\end{document}